\documentclass[12pt,dvips]{article}
\usepackage[dvips]{epsfig}

\def\const{\mbox{const}}
\def\e{{\rm e}}
\def\al{\alpha}
\def\eps{\epsilon}
\def\d{\partial}
\def\l{\left(}
\def\r{\right)}
\def\la{\langle }
\def\ra{\rangle }
\newcommand{\be}{\begin{equation}}
\newcommand{\ee}{\end{equation}}
\newcommand{\tg}{\mathop{\rm tg}\nolimits}
\renewcommand{\ln}{\mathop{\rm ln}\nolimits}
\newcommand{\sm}[1]{{\scriptscriptstyle \rm #1}}
\newcommand{\Tr}{{\rm Tr}}
\newcommand{\bg}{\begin{gather}}
\newcommand{\eg}{\end{gather}}
\def\half{\frac{1}{2}}
\newcommand{\eq}[1]{(\ref{#1})}
\newcommand{\goe}{\gtrsim}
\newcommand{\loe}{\lesssim}

\begin{document}

\title{
Tunneling in quantum cosmology: numerical study of 
particle creation.
}
\author{D.~Levkov$^a$\footnote{levkov@ms2.inr.ac.ru}, 
C.~Rebbi$^b$\footnote{rebbi@bu.edu}, 
V.~A.~Rubakov$^a$\footnote{rubakov@ms2.inr.ac.ru} \\
$^a${\small  Institute for Nuclear Research
  of the Russian Academy of
  Sciences,}\\
{\small 60th October Anniversary Prospect 7a,}\\
{\small Moscow, 117312, Russian Federation}\\
$^b${\small    Boston University, Physics Department,}\\
{\small    590 Commonwealth Avenue,}\\
{\small    Boston, MA 02215, USA}
}

\maketitle          

\begin{abstract}
      We consider a minisuperspace model for a
      closed universe with small and positive cosmological constant
      $\Lambda$,
      filled with a massive scalar field conformally coupled
      to gravity. In the quantum version of this model, the universe
      may undergo a tunneling transition through an effective barrier 
      between regions of small and large scale factor.
      We solve numerically the minisuperspace 
      Wheeler--De~Witt equation with
      tunneling boundary conditions for the wave function of the
      universe, and find that tunneling in quantum cosmology is quite
      different from that in quantum mechanics. Namely, 
      the matter degree of freedom gets 
      excited under the barrier, provided its interaction
      with the scale factor is not too weak, and makes a strong back
      reaction onto tunneling.
      In the  semiclassical limit of small $\Lambda$, 
      the matter energy behind the barrier is close to the height 
      of the barrier: the system ``climbs up'' the barrier, and then
      evolves classically from its top. These features are 
      even more pronounced for 
      inhomogeneous modes of matter field.
      Extrapolating to field theory  we thus argue that high momentum 
      particles are copiously created in the tunneling process.
      Nevertheless,
      we find empirical evidence for the semiclassical-type
      scaling with $\Lambda$
      of the wave function under and behind the barrier. 
\end{abstract}

\newpage
\section{Introduction and summary.}      

A traditional formalism in quantum cosmology is based on the Wheeler--De~Witt
equation~\cite{Wheeler,DeWitt}, often truncated to minisuperspace.
Though this equation is not applicable at length scales near and below
the Planck scale, it should correctly describe quantum gravity
phenomena at larger distances. Among these phenomena, penetration
through classically forbidden regions is of particular interest.
Indeed, various proposals for 
``quantum creation of the 
Universe''~\cite{Vilenkin1,Grish,Linde,Rubakov,Vilenkin2}
invoke this possible quantum gravity effect  in one or another way
(for discussion and references see, e.g.,
Refs.~\cite{Vilenkin3,Rubakov2}).

A prototype example is an empty closed Universe with small and
positive cosmological constant $\Lambda$. In the minisuperspace
approximation, the only dynamical degree of freedom is the radius of
the Universe $a$. In Planck units and with appropriate rescaling,
the gravitational action has the following form (see, e.g.,
Refs.~\cite{Vilenkin3,Rubakov2} and references therein),
\[
 S_a = \int d\eta (\dot{a} \pi_a - N {\cal H}_a)\;,
\]
where $\eta$ is the time variable (in $N=1$ gauge, $\eta$
has the meaning of conformal time), 
$\pi_a$ is momentum conjugate to $a$, 
$N$ is the lapse function and
\[
  {\cal H}_a = -\frac{\pi_a^2}{2} - \left(\frac{a^2}{2} 
-\Lambda a^4 \right)
\]
is the (conformal) Hamiltonian for the scale factor.
Correspondingly, the minisuperspace Wheeler--De~Witt equation is
\begin{equation}
  \hat{{\cal H}}_a \Psi = 0\;,
\label{2*}
\end{equation}
where $\Psi$ is the wave function of the Universe, and $\hat{a}$ and 
$\hat{\pi}_a$ are treated as operators with standard commutation
relations\footnote{The operator ordering ambiguity, inherent in
eq.~(\ref{2*}), is unimportant for our purposes.}. In the coordinate
representation one has $\Psi = \Psi(a)$ and 
$\hat{\pi}_a = -i \partial / \partial a$.
One may modify eq.~(\ref{2*}) slightly by adding a spatially
homogeneous mode of a massless conformal scalar field. Then the
Wheeler--De~Witt equation becomes
\be
    \hat{{\cal H}}_a \Psi  + \epsilon \Psi = 0\;,
\label{2+}
\ee
where $\epsilon$ is the conformal energy of the scalar mode.

\begin{figure}[t]
\begingroup%
  \makeatletter%
  \newcommand{\GNUPLOTspecial}{%
    \@sanitize\catcode`\%=14\relax\special}%
  \setlength{\unitlength}{0.1bp}%
\begin{picture}(3600,2160)(0,0)%
\special{psfile=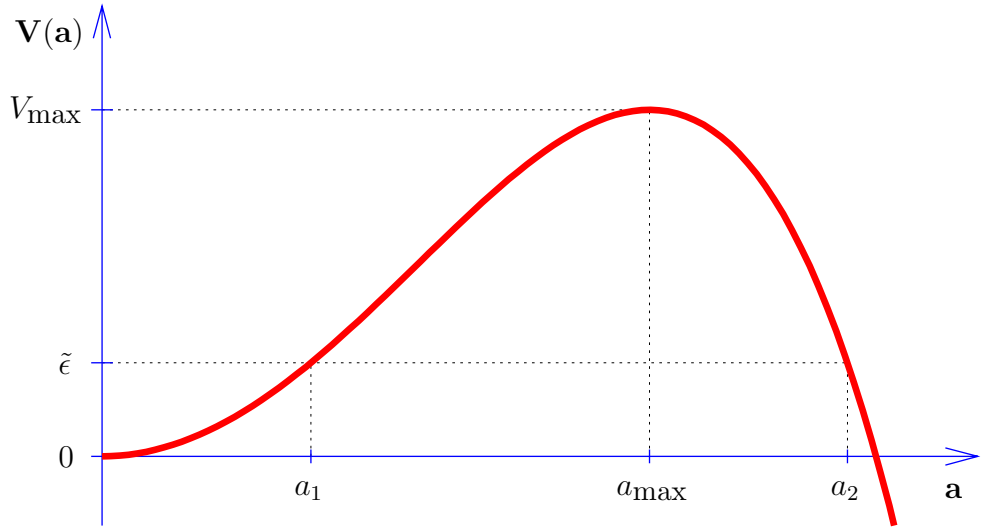 llx=0 lly=0 urx=720 ury=504 rwi=7200}
\put(3326,231){\makebox(0,0)[l]{$\mathbf a$}}%
\put(2897,231){\makebox(0,0)[l]{$a_2$}}%
\put(2089,231){\makebox(0,0)[l]{$a_{\mbox{\small max}}$}}%
\put(875,231){\makebox(0,0)[l]{$a_1$}}%
\put(-14,361){\makebox(0,0)[l]{$0$}}%
\put(-14,714){\makebox(0,0)[l]{$\tilde\epsilon$}}%
\put(-200,1668){\makebox(0,0)[l]{$V_{\mbox{\small max}}$}}%
\put(-179,1955){\makebox(0,0)[l]{$\mathbf{V(a)}$}}%
\end{picture}%
\endgroup
 
\caption{\label{fig21} Potential barrier in eq.~(\ref{2+}).}
\end{figure}

There are various proposals for the wave function of the
Universe~{\cite{Vilenkin1,HartleHawking,Linde,Andy}} which, in the language
of the Wheeler--De~Witt equation, correspond to different boundary
conditions for $\Psi(a)$. In this paper we consider tunneling
boundary conditions~\cite{Vilenkin1,Rubakov,Vilenkin2}. These 
resemble closely 
the boundary conditions corresponding to tunneling transitions in
conventional quantum mechanics. Namely, eq.~(\ref{2+}) has the form of
the stationary Schr\"odinger equation at energy $\epsilon$ in a
potential
\[
   V(a) = \frac{a^2}{2} - \Lambda a^4\;,
\]
see Fig.~\ref{fig21}. At small $\Lambda$ (in Planck units)
the system is semiclassical: it is straightforward to see that
$\Lambda$ plays the role of $\hbar$. At sufficiently 
large~$\epsilon$ (again in Planck units), but 
$\epsilon < V_{max} = 1/(16 \Lambda)$, there are classically allowed
regions on the two sides of the barrier. In conventional quantum
mechanics, tunneling from small $a$ to large $a$ would be described by
the wave function which contains only outgoing waves as $a \to
+\infty$; if there are degrees of freedom other than $a$, then the
incoming component of the wave function has to be specified on
the left of the barrier. In this paper we consider the same boundary
conditions in the context of quantum cosmology.

At small $\Lambda$, the decaying part of the wave function in the
classically forbidden region is
\be
\Psi(a) \propto \mbox{e}^{- \displaystyle\int da \sqrt{2 (V(a) - \epsilon)}}\;,
\label{4+}
\ee
so the wave function behind the barrier 
is suppressed by the standard semiclassical exponential
\be
    \left.\Psi \right|_{a\to+\infty} \sim 
\mbox{e}^{- \displaystyle\int\limits_{a_1}^{a_2}~da~\sqrt{2 (V(a) - \epsilon)}}
\label{4*}
\ee
(assuming that $\Psi \sim 1 $ in the classically allowed region on the
left of the barrier), where $a_1$ and $a_2$ are the two turning
points.

The analogy to the stationary Schr\"odinger equation is no longer
perfect if matter degrees of freedom (or gravitons) are
added. Consider, as an example, a massive scalar field conformally
coupled to gravity, whose action is
\[
  S_{\Phi} = \int~d^4x~\sqrt{-g}\left(\half g^{\mu \nu} 
\partial_\mu \Phi \partial_\nu \Phi - \half m^2 \Phi^2
+ \frac{1}{12} R \Phi^2 \right)\;.
\]
In the minisuperspace approximation, the scalar field is spatially
homogeneous,
\[
    \Phi =  \frac{\phi(\eta)}{a(\eta)}\;,
\]
and its action, with appropriate rescaling, is
\[
  S_{\phi} = \int~d\eta~(\dot{\phi} \pi_\phi - N {\cal H}_{\phi})\;,
\]
where
\be
    {\cal H}_\phi = \frac{\pi_\phi^2}{2} +
\half (1 + m^2 a^2) \phi^2\;.
\label{4**}
\ee
We note in passing that if one blindly considered 
an inhomogeneous mode of conformal momentum $k$, then the scalar
Hamiltonian would be
 \be
    {\cal H}_\phi = \frac{\pi_\phi^2}{2} +
\half (k^2 + 1 + m^2 a^2) \phi^2\;.
\label{5+}
\ee
With the scalar field added, the Wheeler--De~Witt equation becomes
\be
  (\hat{{\cal H}}_a + \hat{{\cal H}}_\phi) \Psi + \epsilon \Psi = 0\;.
\label{5*}
\ee
Unlike in conventional quantum mechanics,  this equation contains
a negative ``kinetic term'' for the scale factor, $(-\pi_a^2 /2)$,
and the potential $V(a)$ enters with negative sign also.
This feature, which at first sight appears technical (and is often
ignored in literature, see, however, Ref.~\cite{Kamenshchik:1995ib}),
in fact reflects the difference between tunneling in quantum cosmology
and in quantum mechanics. Consider, as an example, tunneling from the 
ground state of the $\phi$-oscillator.
In quantum mechanics, the excitation of a
fluctuating degree of freedom (similar to
the field $\phi$) would lower the
energy of the tunneling degree of freedom (the scale factor $a$), and
thus suppress tunneling even further. Hence, the fluctuating degree of
freedom does not get excited much and the semiclassical exponent
(\ref{4*}) does not change (the fluctuating degree of freedom affects
the pre-exponential factor only). In quantum cosmology, on the
contrary, the excitation of the field $\phi$ makes the tunneling
of the Universe exponentially easier~\cite{Rubakov}: crudely speaking,
the effective value of $\epsilon$ gets increased. 
Hence, the $\phi$-quanta tend to be copiously created. 
Because of their back reaction,
the semiclassical
expansion about the solution (\ref{4+}) may break down, and the entire
picture of tunneling in quantum cosmology may get substantially modified
 when matter degrees of freedom or gravitons are
added. 

It is worth pointing out, that at given $\epsilon$
there exists only one real Euclidean
solution in the theory with the action $(S_a+S_\phi)$. This solution,
the instanton (periodic instanton for $\epsilon \neq 0$),  has 
$\phi = 0$ and coincides with the instanton  in a theory
truncated to a single variable $a$. The above arguments imply that
this instanton  may become completely irrelevant if matter degrees 
of freedom and/or gravitons are kept.

In this paper we substantiate our qualitative observations by a
numerical study of quantum systems of the Wheeler--De~Witt
type. Namely, we consider systems with two degrees of freedom whose
own
Hamiltonians are of opposite
signs. To obtain a clear picture, we modify eq.~(\ref{5*}) in such a
way that the potential $V(a)$ is well localized in $a$, and the
interaction between~$\phi$ and $a$ is also localized in $a$. Then
the tunneling boundary conditions are well defined even for a finite
interval of $a$ (which one has to work with in a numerical study), while
the peculiarities of the Wheeler--De~Witt equation remain there.
To verify that the main results of our study are not sensitive to 
the details of the $a$-$\phi$ interaction, we modify the interaction 
in different ways, and consider three different models in subsequent 
sections. Our particular systems are described in section 2.1.

If the interaction between $\phi$ and $a$ is weak enough, the matter
degree of freedom may be treated as perturbation (provided it is in
its ground state on the left of the barrier, otherwise the parameter
$\epsilon$ should be redefined). To the zeroth approximation, the wave
function in the classically forbidden region is then given by
eq.~(\ref{4+}). In section 2.2 we present a perturbative treatment of
our systems about the solution (\ref{4+}) and get a heuristic
understanding of the range of parameters in which the perturbation
theory breaks down. Unlike in conventional quantum mechanics, this
range includes part of the would-be semiclassical region of small
$\Lambda$. We will be mostly interested in this range of parameters.

Our main purpose is to study our systems at the quantum level.
We have solved the Wheeler--De~Witt equation for these systems
numerically, without making use of any approximation (except for
approximations inevitable in numerical studies, such as
discretization, truncation of the matter Hamiltonian to 
finite number of levels, etc.; these approximations are safely under
control). We describe our results in section 3. We begin with the
numerical analysis of the classical evolution (section 3.1) to see
that, in a wide range of parameters, the potential barrier cannot be
overcome classically. We then proceed in section 3.2
to the quantum treatment. We find
that unless the interaction between $a$ and $\phi$ is weak, the matter
degree of freedom does indeed get strongly excited under the barrier, and
the system behind the barrier contains a large number of $\phi$-quanta.
Furthermore, in the region of the parameter space which in
conventional quantum mechanics would be adequately described 
semiclassically, with the wave function (\ref{4+}), the energy
of the matter degree of freedom is close, behind the barrier, to
the height of the barrier $V_{max} \sim 1/\Lambda$. 
The system ``climbs up'' the
potential barrier, and the classically allowed 
region effectively starts
again just right of the point of maximum, $a_{max}$. These features
are even more pronounced for a {\it larger} $a$-independent term in the
oscillator frequency, $k^2$ in eq.~(\ref{5+}), again in contrast to
quantum mechanics.
Even if the interaction between matter and gravity is weak (i.e., the
parameter $m$ in eq.~\eq{5+} is small), the $\phi$-quanta are
copiously created due to tunneling, provided that their frequency is
high ($k$ is large enough).
 Extrapolating to field theory in the tunneling
Universe, we thus argue that high momentum modes get strongly excited
in the tunneling process.

In section 3.3 we concentrate on scaling properties in $\Lambda$
of our quantum
solutions in the region of  small $\Lambda$. We find, somewhat
surprisingly, that the wave functions in fact scale in a semiclassical
way, $\ln \Psi =  {\cal F}/\Lambda$, where ${\cal F}$ 
depends on appropriate
combinations of other parameters and $\Lambda$. This empirical
evidence suggests that tunneling in quantum cosmology may still be
described semiclassically, but the relevant classical solutions are
entirely different from the instanton existing in this model.
This conjecture is further supported by our study
of the dependence on the $a$-$\phi$ coupling parameter ($m$ in
eq.~(\ref{4**})) at small but fixed $\Lambda$. As we already mentioned,
at small $m$ the matter
degree of freedom is not excited due to tunneling, while at larger $m$
it is, and its energy behind the barrier is close to $V_{max}$. We
find numerically that the transition between the two regimes is very
sharp in $m$ and becomes close to step function as $\Lambda$ gets 
smaller.
 This suggests that the wave function is a combination of
two semiclassical exponentials,
\be
    \Psi = C_1 \mbox{e}^{-S_1} +  C_2 \mbox{e}^{-S_2}
\label{10*}
\ee
where the first exponential is due to tunneling along $\phi =0$ 
and coincides with eq.~(\ref{4+}), while the second exponential 
comes from an
entirely different classical path. In spite of considerable effort, we
were unable to identify the classical solution responsible for
tunneling with excitation of the matter degree of freedom,
which would be of considerable interest to quantum cosmology.

Overall, we have found that tunneling in quantum cosmology is quite
different from that in quantum mechanics. This observation may have
interesting implications to the discussion of initial conditions for
the classical stage of the cosmological evolution.

Appendix A contains details of our numerical procedure. We describe
various checks of this procedure in Appendix B. These checks ensure
that our results are reliable, even though we are dealing with very
small numbers typical for tunneling. The calculations have been  
performed at the supercomputer SGI Origin 2000 of the Center for
Computational Science of Boston University.

\section{The model and perturbative solution.}

It is convenient to rescale the variable $a$ by introducing
\[
        b = \sqrt{\Lambda} a
\]
and write the Wheeler--De~Witt equation (\ref{5*}) in the following
form,
\be
\left[ \half \Lambda \frac{\partial^2}{\partial b^2}
- \frac{1}{\Lambda} \left( U(b) - \tilde{\epsilon} \right)
+ \hat{{\cal H}}_\phi \right] \Psi (b) = 0\;,
\label{11*}
\ee
where the coordinate representation in $b$ is chosen, while the
representation of operators $\hat{\pi}_\phi$ and $\hat{\phi}$ is not
yet fixed. Here $\hat{{\cal H}}_\phi$ is the Hamiltonian of an
oscillator with $b$-dependent frequency,
\be
 \hat{{\cal H}}_\phi = \half \hat{\pi}_\phi^2 +
\half \Omega^2 (b) \hat{\phi}^2
\label{11+}
\ee
where
\be
  \Omega^2 (b) = \omega^2 + M^2 f(b)
\label{11**}
\ee
and we introduced a parameter $\omega$ for 
generality\footnote{For inhomogeneous modes, one would have
$\omega^2 = 1 + k^2$, see eq.~\eq{5+}. Note that $k$ 
is the conformal momentum of
the scalar mode, which is related to the physical momentum
as $p_{phys} = k/a = k\sqrt{\Lambda} /b$. 
Even for modes with sub-Planckian
physical momentum, the value of $k$, and hence $\omega$, may be much 
larger than 1.}. 
The parameters~$\tilde{\epsilon}$ and $M$ are related
to original parameters as follows,
\be
\tilde{\epsilon} = \Lambda \epsilon \;, \;\;\;\;\;
M^2 = \frac{m^2}{\Lambda}\;.
\label{add1}
\ee
It is worth noting that for arbitrary
$U(b)$ and $f(b)$, eq.~(\ref{11*}) corresponds to the action
\begin{eqnarray}
  S = &-& \frac{1}{\Lambda} \int~d\eta~ \left[ \frac{\dot{b}^2}{2N}
- N \left( U(b) - \tilde{\epsilon}\right) \right] 
\nonumber \\
&+& \int~d\eta~\left[ \frac{\dot{\phi}^2}{2N} - \half
N \Omega^2(b) \phi^2 \right]\;,
\label{12*}
\end{eqnarray}
where $N$ is the lapse function.

In the original Wheeler--De~Witt equation, one has
\[
U(b) = \Lambda V(b/\sqrt{\Lambda}) = (b^2/2 - b^4)
\]
and $f(b) = b^2$. In our numerical study, however, we found it
convenient to choose $U(b)$ and $f(b)$ is such a way that they are
well localized.

\subsection{Modified models.}
We note that, since we are imposing {\it free} boundary conditions
at $b = 0$ (as $U(0) = 0$,  $f(0) = 0$), the range of $b$ can be 
extended to $(-\infty ;\;\;+\infty)$.
It is then convenient to shift $b$ in such a way that the maximum of the 
potential  $U(b)$ occurs at $b = 0$.
To be able to
impose the boundary conditions in a finite interval of $b$, we will
work with the following potential and interaction functions,
\be
  U(b) = \mbox{e}^{-\displaystyle{b^2}/2}
\label{12**}
\ee
and
\[
f(b) = \displaystyle\frac{c}{\mbox{e}^{\alpha_1 (b - b_1)} + 
\mbox{e}^{\alpha_2(b - b_2)} 
+ \mbox{e}^{\alpha_3 (b - b_3)}} \label{f}. 
\]
We have chosen three sets of parameters $\alpha_{1}$,
$\alpha_2$, $\alpha_3$ and $b_1$, $b_2$, $b_3$, summarized in Table 1. 
Since not all $\alpha_i$ have the same sign, the functions
$f_1$, $f_2$ and $f_3$ all decay exponentially as $|b| \to \infty$.
We have chosen values of $c$ such that the maximum values of these
functions are close to 1. Then the strength of $b$-$\phi$ interaction
is determined by the parameter $M$ in eq.~(\ref{11**}).

\begin{table}
\label{Table:1}
\begin{center}
\begin{tabular}{|c|c|c|c|c|c|c|c|}
\hline
 & $c$ & $\alpha_1$ & $b_1$ & $\alpha_2$ & $b_2$ & $\alpha_3$ & $b_3$\\
\hline
$f_1$ & $1.356$ & $0.2$ & $- 6$ & $- 2$ & $- 6$ & $3$ & $5$\\
\hline
$f_2$ & $1.356$ & $- 0.2$ & $6$ & $2$ & $6$ &  $- 3$ & 0\\
\hline
$f_3$ & $1.454$ & $- 0.225$ & $0.7$ & $3$ & $0.7$ & $- 2.55$ & $- 0.65$\\
\hline
\end{tabular}
\end{center}
\caption{Parameters of the interaction functions.}
\end{table}

\begin{figure}[ht!]
\begingroup%
  \makeatletter%
  \newcommand{\GNUPLOTspecial}{%
    \@sanitize\catcode`\%=14\relax\special}%
  \setlength{\unitlength}{0.1bp}%
\begin{picture}(3600,2160)(0,0)%
\special{psfile=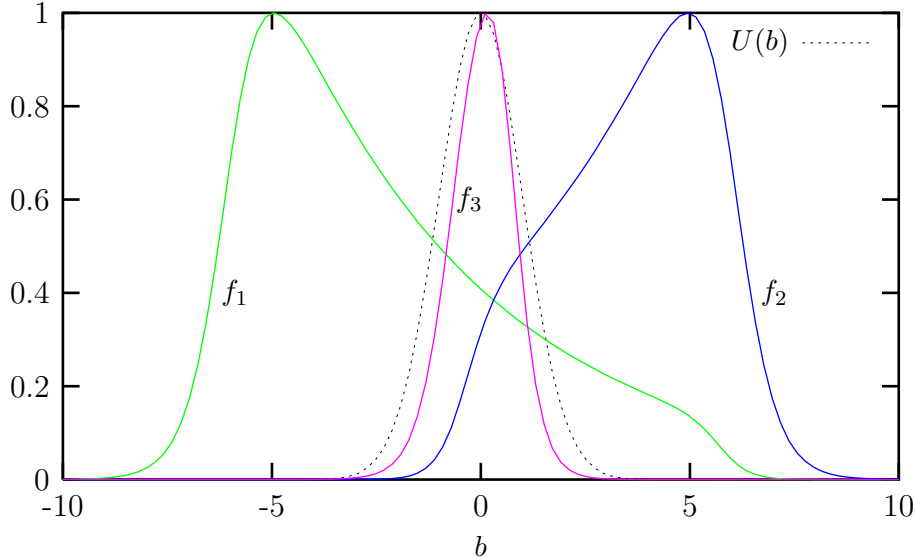 llx=0 lly=0 urx=720 ury=504 rwi=7200}
\put(3037,1947){\makebox(0,0)[r]{\small $U(b)$}}%
\put(1781,1356){\makebox(0,0)[l]{\small $f_3$}}%
\put(2930,1004){\makebox(0,0)[l]{\small $f_2$}}%
\put(898,1004){\makebox(0,0)[l]{\small $f_1$}}%
\put(1875,50){\makebox(0,0){\small $b$}}%
\put(3450,200){\makebox(0,0){10}}%
\put(2663,200){\makebox(0,0){5}}%
\put(1875,200){\makebox(0,0){0}}%
\put(1088,200){\makebox(0,0){-5}}%
\put(300,200){\makebox(0,0){-10}}%
\put(250,2060){\makebox(0,0)[r]{1}}%
\put(250,1708){\makebox(0,0)[r]{0.8}}%
\put(250,1356){\makebox(0,0)[r]{0.6}}%
\put(250,1004){\makebox(0,0)[r]{0.4}}%
\put(250,652){\makebox(0,0)[r]{0.2}}%
\put(250,300){\makebox(0,0)[r]{0}}%
\end{picture}%
\endgroup
 
\caption{\label{fig0}
The potential $U(b)$ and the three interaction functions.}
\end{figure}

The potential $U(b)$ and the three interaction functions 
are shown in Fig.~\ref{fig0}. The shapes of the interaction functions are
quite different: while $f_1$ is substantial in front of the barrier
(and hence the matter degree of reedom $\phi$ may be excited there),
$f_2$ and $f_3$ vanish in front of the barrier, with $f_3$ being
non-zero in the classically forbidden region only. Thus, these
three interaction functions 
constitute a representative set. We will see that the main features of 
tunneling do not depend on the shape of the interaction function.

For fixed choice of  $f(b)$, there are four parameters in the model,
namely, $\Lambda$, $\tilde{\epsilon}$, $\omega$ and $M$. Naively, one
would expect either from eq.~(\ref{11*}) or eq.~(\ref{12*}) that as
$\Lambda \to 0$ with $\tilde{\epsilon}$, $\omega$ and $M$ fixed,
tunneling through the barrier $U(b)$ is described semiclassically,
and, furthermore, that the matter degree of freedom makes a small effect
(provided that tunneling occurs from a low-lying state of the
$\phi$-oscillator). Let us consider the corresponding perturbation
theory to see whether this is indeed the case.

\subsection{Perturbative treatment.}

Let us consider the case $\Lambda \ll 1$ with $\tilde{\epsilon}$,
$\omega$ and $M$ fixed, and treat $\hat{{\cal H}}_\phi$ in
eq.~(\ref{11*}) as perturbation. We take the $\phi$-oscillator at left of
the barrier in its ground state. Neglecting $\hat{{\cal H}}_\phi$
altogther, we obtain that in the classically forbidden region, except
near the turning points, the dominant part of the wave function is
\be
   \Psi_0 (b) = \frac{\mbox{e}^{- S_0 (b)}}
{[2(U(b) - \tilde{\epsilon})]^{1/4}}\;,
\label{14*}
\ee
where
\[
S_0 (b) = \frac{1}{\Lambda} \int_{b_1}^{b}~db~\sqrt{2(U(b) -
\tilde{\epsilon})} 
\]
and $b_1$ is the left turning point. To include the effect of
$\hat{{\cal H}}_\phi$ perturbatively, we write
\[
  \Psi (b) = \frac{\mbox{e}^{-S_0 (b)}}{[2(U(b) - \tilde{\epsilon})]^{1/4}} 
 \Psi_{\phi} (b)\;,
\]
and obtain, to the first non-trivial order, that $\Psi_\phi (b)$ obeys
\be
    \frac{d \Psi_\phi}{d \tau} =
 + \hat{{\cal H}}_\phi (\tau) \Psi_\phi\;,
\label{15*}
\ee
where we introduced a variable $\tau$, instead of $b$, such that
\[
  \frac{db}{d\tau} = \sqrt{2(U(b) - \tilde{\epsilon})}
\]
and $\tau = 0$ corresponds to the left turning point.
Equation (\ref{15*}) is a ``wrong sign'' Euclidean Schr\"odinger
equation; the ``wrong sign'' (as compared to conventional quantum
mechanics) originates from the ``wrong sign'' of the Hamiltonian for
the scale factor~\cite{Rubakov}.

To deal with eq.~(\ref{15*}), let us decompose $\Psi_\phi (\tau)$ into
eigenstates of the ``instantaneous'' Hamiltonian \eq{11+}, i.e., the
eigenstates $\Psi_n^{(\Omega)} (\tau)$ of an oscillator of frequency 
$\Omega(\tau) \equiv \Omega(b(\tau))$,
\[
   \Psi_\phi (\tau) = \sum_{n=0}^{\infty}
c_n (\tau) \Psi_n^{(\Omega)}   (\tau)\;.
\]
We find that \eq{15*} translates into a set of equaions for the
coefficients $c_n$,
\be
 \frac{d c_n}{d \tau} =  \Omega (n + 1/2) c_n 
- \frac{1}{4 \Omega} \frac{d \Omega}{d \tau}
\sqrt{(n + 1)(n + 2)} c_{n + 2} +
\frac{1}{4 \Omega} \frac{d \Omega}{d \tau} \sqrt{n (n - 1)} c_{n - 2}\;.
\label{16*}
\ee
This set is still hard to solve. To get an idea of its solutions, we
make use of the adiabatic approximation, and treat the last two terms 
in
eq.~\eq{16*} perturbatively. Since we consider an oscillator in its ground
state on the left of the barrier, the zeroth order solution is
\begin{equation}
  c_n^{(0)} (\tau) = \mbox{e}^{\half\int\limits_0^\tau d\tau' 
\Omega(\tau')} \delta_{n,\;0} \;.
\end{equation}
To the first order we obtain that the only non-vanishing coefficient
is
\be
  c_2^{(1)} (\tau) =  
\mbox{e}^{\half\int\limits_0^\tau d\tau' \Omega(\tau')}
\int\limits_0^\tau d \tau' \frac{1}{2 \sqrt{2}
\Omega(\tau')}
  \frac{d\Omega(\tau')}{d \tau'} 
\mbox{e}^{2 \int\limits_{\tau'}^\tau \Omega(\tau^{\prime\prime})
d\tau^{\prime\prime} }\;,
\label{16+}
\ee
The first factor in the integrand,
\[
 \frac{1}{2 \sqrt{2}\Omega(\tau')}
  \frac{d\Omega(\tau')}{d \tau'} 
\equiv  \frac{1}{2 \Omega(b')}
  \frac{d\Omega(b')}{d b'} \sqrt{U(b') - \tilde{\epsilon}}
\]
accounts for the creation of a pair of $\phi$-quanta at point $b'$ in
the forbidden region; it is small in the adiabatic regime,
 \[
 \frac{1}{\Omega(\tau')}
  \frac{d\Omega(\tau')}{d \tau'} \ll 1\;,
\]
which occurs at large $\omega$ or small $M$. However, there is a
growing exponential factor in the integrand of eq.~\eq{16+}, which
reflects the property that the creation of $\phi$-quanta enhances
tunneling. The perturbation theory about the solution \eq{14*}
breaks down when this factor becomes dominant; in that case one
expects that matter has strong back reaction on the  process
of tunneling.

Our perturbative treatment shows that at given $\omega$
 and small $M$, tunneling in quantum cosmology is indeed described by
 the wave function \eq{14*} and effects of the matter degree of
 freedom are small. However, as $M$ increases,
the tunneling regime changes, and
matter effects become important. A similar phenomenon occurs if $M$
 is kept fixed and small, while $\omega$ increases instead: high
 energy quanta are copiously created even if their interaction
to gravity is weak\footnote{According to eq.~\eq{add1}
this may happen even if the physical masses and momenta are small.}. 
This is in sharp contrast to
 conventional quantum mechanics.

\section{Numerical analysis.}

\subsection{Classical solutions.}

The system with the action \eq{12*} may either tunnel through the
potential barrier or overcome it classically, depending on the
parameters. Before studying tunneling, it is instructive to discuss
briefly the classical solutions and find the range of parameters in 
which classical transitions over the barrier do not occur.

Let the classical system begin its evolution from large negative $b$
towards the barrier. In terms of the variables $b$ and
\[
\varphi = \sqrt{\Lambda} \phi
\]
the classical equations of motion read
\begin{eqnarray}
\frac{d^2 b}{dt^2} &=& - \frac{dU(b)}{db} + \frac{1}{2} M^2 \varphi^2 \frac{df(b)}{db} 
\nonumber\;, \\
\frac{d^2 \varphi}{dt^2} &=& - \Omega^2(b) \varphi\;.
\label{19*}
\end{eqnarray}
Here we fixed the gauge by introducing the standard cosmic time 
$t = \int\limits^\eta d\eta' N(\eta')$.
The classical constraint 
\be
  {\cal H}_a + {\cal H}_\phi + \tilde{\epsilon} = 0
\label{20*}
\ee
should also be satisfied.

Note that the parameter
$\Lambda$ does not enter the classical 
equations; as we pointed out above,
it is analogous to $\hbar$ and appears in the quantum problem only.

At large negative $b$ the variables decouple. One can freely choose
 the
 value of
$b$ in this region at initial time. Then there are two independent
 initial data, which we choose as the $\phi$-oscillator energy 
$E_\varphi (t=0) \equiv E_{\varphi, 0}$ and its initial phase
 $\theta_0$.  The ``energy'' associated with the scale factor,
which we define as
\[
  E_b = - {\cal H}_a\;,
\]
is then determined at the initial moment of time from the constraint
\eq{20*}, 
\be
  E_{b,0} = E_{\varphi,0} + \frac{\tilde{\epsilon}}{\Lambda}\;.
\label{20**}
\ee
For given initial $b$ the latter relation determines $\dot{b}(t=0)
\equiv - \pi_b (t=0)$, so the initial data are completely defined.
Conversely, one may choose the initial value of $\pi_b$, and hence
$E_{b,0}$, arbitrarily, and find $\tilde{\epsilon}$ from
eq.~\eq{20**}. Note that
\be
      E_{b,0} \geq E_{\varphi,0}
\label{20+}
\ee
since $\tilde{\epsilon}$ is non-negative.

For every pair of parameters $(M,\omega)$ there exists a region in the
plane $(E_{\varphi,0}; E_{b,0})$ in which at some values of the phase
$\theta_0$ classical over-barrier transitions occur. The rest of
this plane may be called ``the no-transition region'' of initial data:
irrespectively of the oscillator phase, the barrier cannot be overcome
classically for initial data belonging to this region. Let us note
that at $E_{\varphi,0} = 0$ the $\phi$-oscillator remains in its
classical ground state $\varphi =0$ during the entire evolution, so
the initial data {$(E_{\varphi,0} =0; E_{b,0} > U_{max}/\Lambda \equiv
1/\Lambda )$} give rise to classically allowed over-barrier transitions,
while the interval $(E_{\varphi,0} =0; 0< E_{b,0} < 1/\Lambda )$
belongs to the no-transition region of initial data.

For $E_{\varphi,0} \neq 0$  classical over-barrier transitions may
occur even if $E_{b,0}$ is lower than the barrier height
$1/\Lambda$. Indeed the parametric amplification of
$\varphi$-oscillations may take place and give rise to over-barrier
transitions for $E_{b,0} < 1/\Lambda$. Hence, the boundary of the
no-transition region of initial data is not a straight line; this
boundary depends on parameters $M$ and $\omega$. 

\begin{figure}[ht!]
\begingroup%
  \makeatletter%
  \newcommand{\GNUPLOTspecial}{%
    \@sanitize\catcode`\%=14\relax\special}%
  \setlength{\unitlength}{0.1bp}%
\begin{picture}(3600,2160)(0,0)%
\special{psfile=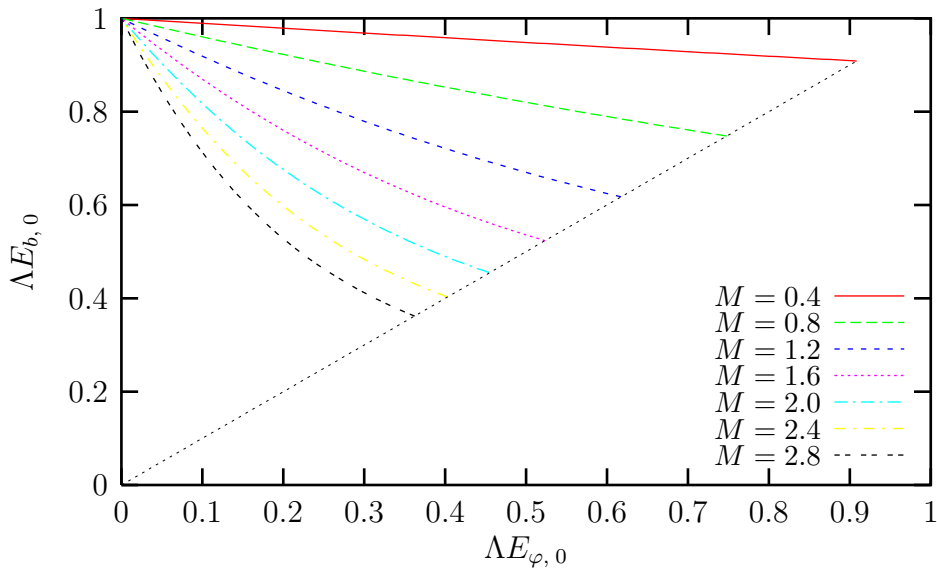 llx=0 lly=0 urx=720 ury=504 rwi=7200}
\put(3037,413){\makebox(0,0)[r]{\small ${M = 2.8}$}}%
\put(3037,513){\makebox(0,0)[r]{\small ${M = 2.4}$}}%
\put(3037,613){\makebox(0,0)[r]{\small ${M = 2.0}$}}%
\put(3037,713){\makebox(0,0)[r]{\small ${M = 1.6}$}}%
\put(3037,813){\makebox(0,0)[r]{\small ${M = 1.2}$}}%
\put(3037,913){\makebox(0,0)[r]{\small ${M = 0.8}$}}%
\put(3037,1013){\makebox(0,0)[r]{\small ${M = 0.4}$}}%
\put(1925,50){\makebox(0,0){$\Lambda E_{\varphi,\;0}$}}%
\put(100,1180){%
\special{ps: gsave currentpoint currentpoint translate
270 rotate neg exch neg exch translate}%
\makebox(0,0)[b]{\shortstack{$\Lambda E_{b,\; 0}$}}%
\special{ps: currentpoint grestore moveto}%
}%
\put(3450,200){\makebox(0,0){1}}%
\put(3145,200){\makebox(0,0){0.9}}%
\put(2840,200){\makebox(0,0){0.8}}%
\put(2535,200){\makebox(0,0){0.7}}%
\put(2230,200){\makebox(0,0){0.6}}%
\put(1925,200){\makebox(0,0){0.5}}%
\put(1620,200){\makebox(0,0){0.4}}%
\put(1315,200){\makebox(0,0){0.3}}%
\put(1010,200){\makebox(0,0){0.2}}%
\put(705,200){\makebox(0,0){0.1}}%
\put(400,200){\makebox(0,0){0}}%
\put(350,2060){\makebox(0,0)[r]{1}}%
\put(350,1708){\makebox(0,0)[r]{0.8}}%
\put(350,1356){\makebox(0,0)[r]{0.6}}%
\put(350,1004){\makebox(0,0)[r]{0.4}}%
\put(350,652){\makebox(0,0)[r]{0.2}}%
\put(350,300){\makebox(0,0)[r]{0}}%
\end{picture}%
\endgroup
 
\caption{\label{fig4} Boundaries of no-transition regions in the
plane of initial data for $\omega = 0.6$ and different $M$. 
The interaction function is $f_2$. 
Classical over-barrier transitions do not occur for initial data below
these lines and any initial phase of the $\phi$-oscillator.
The dotted line corresponds to $E_{b,0} = E_{\varphi, 0}$, see eq.~\eq{20+}.}
\end{figure}

We have found the
regions in the plane of initial data $(E_{\varphi,0}; E_{b,0})$ in 
which classical over-barrier transitions occur at some values of the
initial oscillator phase $\theta_0$, and those where these transitions
do not happen for any $\theta_0$, by solving numerically the system
\eq{19*}, \eq{20*}. Details of the numerical procedure are given in
Appendix A. We have found that even at relatively large interaction
strengh $M$, classical transitions do not occur for $E_{\varphi, 0}
\ll 1/\Lambda$ and $E_{b,0} < 1/\Lambda$. An example is shown in
Fig.~\ref{fig4}. In what follows we always choose the parameters in such a
way that the classically allowed over-barrier transitions are
impossible. 

\subsection{Solutions to the Wheeler--De~Witt equation.}

Our main purpose is to study solutions to the Wheeler--de~Witt 
equation \eq{11*} for our model systems numerically, without
making use of any approximation.
The tunneling boundary conditions are as follows. At large $|b|$ the
variables $b$ and $\phi$ decouple, and the frequency of the
$\phi$-oscillator is equal to $\omega$. Hence, the asymptotic wave
function may be decomposed into the oscillator eigenstates,
\be
  \Psi (b) = \sum_{n=0}^{\infty} C_n^{\pm} (b) \Psi_n^{(\omega)}
\; , \;\;\;\; b \to \pm \infty\;.
\label{23*}
\ee
At $b \to \pm \infty$ the asymptotic solutions are, respectively,
\[
 C_n^{\pm} (b) = t^{\pm}_n \mbox{e}^{ik_n b}
+  r^{\pm}_n \mbox{e}^{-ik_n b}
\]
where
\[
    k_n = \sqrt{\frac{2}{\Lambda}\left[ \frac{\tilde{\epsilon}}{\Lambda} 
+ \omega\left(n +\half \right) \right]}\;.
\]
We consider tunneling from a state
of fixed oscillator excitation number $n_0$. Then the
tunneling boundary conditions are that the negative momentum component
of the wave function vanishes on the right of the barrier, while the
positive momentum component on the left of the barrier has $n=n_0$,
\begin{eqnarray}
   r_n^+ &=& 0\;,\nonumber
\\
  t_n^- &=& \delta_{n, n_0}\;.
\label{24*}
\end{eqnarray}
These conditions are direct analogs of the tunneling boundary
conditions in quantum mechanics.

Unless explicitly stated, we will consider tunneling from the
ground state of the $\phi$-oscillator,
\[
      n_0 = 0\;.
\]
We will discuss the dependence on $n_0$ towards the end of this 
section.

We describe our numerical procedure in Appendix A. Various checks of
this procedure are summarized in Appendix B. Here we present the
results of our calculations. 

\begin{figure}[ht!]
\begingroup%
  \makeatletter%
  \newcommand{\GNUPLOTspecial}{%
    \@sanitize\catcode`\%=14\relax\special}%
  \setlength{\unitlength}{0.1bp}%
\begin{picture}(3600,2160)(0,0)%
\special{psfile=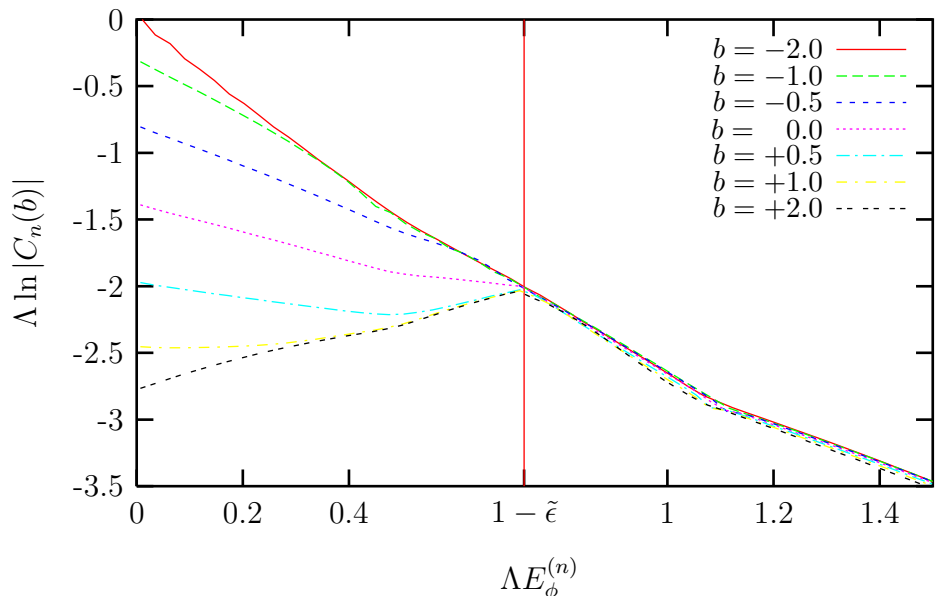 llx=0 lly=0 urx=720 ury=504 rwi=7200}
\put(3037,1347){\makebox(0,0)[r]{\small $b = + 2.0$}}%
\put(3037,1447){\makebox(0,0)[r]{\small $b = + 1.0$}}%
\put(3037,1547){\makebox(0,0)[r]{\small $b = + 0.5$}}%
\put(3037,1647){\makebox(0,0)[r]{\small $b =\;\;\;  0.0$}}%
\put(3037,1747){\makebox(0,0)[r]{\small $b = - 0.5$}}%
\put(3037,1847){\makebox(0,0)[r]{\small $b = - 1.0$}}%
\put(3037,1947){\makebox(0,0)[r]{\small $b = - 2.0$}}%
\put(1950,-50){\makebox(0,0){ $\Lambda E_\phi^{(n)}$}}%
\put(100,1180){%
\special{ps: gsave currentpoint currentpoint translate
270 rotate neg exch neg exch translate}%
\makebox(0,0)[b]{\shortstack{$\Lambda \ln |C_n(b)|$}}%
\special{ps: currentpoint grestore moveto}%
}%
\put(3250,200){\makebox(0,0){1.4}}%
\put(2850,200){\makebox(0,0){$1.2$}}%
\put(2450,200){\makebox(0,0){1}}%
\put(1910,200){\makebox(0,0){$1 - \tilde\epsilon$}}%
\put(1250,200){\makebox(0,0){$0.4$}}%
\put(850,200){\makebox(0,0){$0.2$}}%
\put(450,200){\makebox(0,0){$0$}}%
\put(400,2060){\makebox(0,0)[r]{0}}%
\put(400,1809){\makebox(0,0)[r]{-0.5}}%
\put(400,1557){\makebox(0,0)[r]{-1}}%
\put(400,1306){\makebox(0,0)[r]{-1.5}}%
\put(400,1054){\makebox(0,0)[r]{-2}}%
\put(400,803){\makebox(0,0)[r]{-2.5}}%
\put(400,551){\makebox(0,0)[r]{-3}}%
\put(400,300){\makebox(0,0)[r]{-3.5}}%
\end{picture}%
\endgroup
 
\caption{\label{fig17} The occupation number as function of
the energy of the $\phi$-oscillator at different values of the scale
factor, for the interaction function $f_1$ and $n_0=0$ (tunneling from
the oscillator ground state), $\Lambda = 0.0225$, $M=0.2$,
$\omega=0.6$,
$\tilde{\epsilon}= 0.27$.}
\end{figure}

\begin{figure}[ht!]
\begingroup%
  \makeatletter%
  \newcommand{\GNUPLOTspecial}{%
    \@sanitize\catcode`\%=14\relax\special}%
  \setlength{\unitlength}{0.1bp}%
\begin{picture}(3600,2160)(0,0)%
\special{psfile=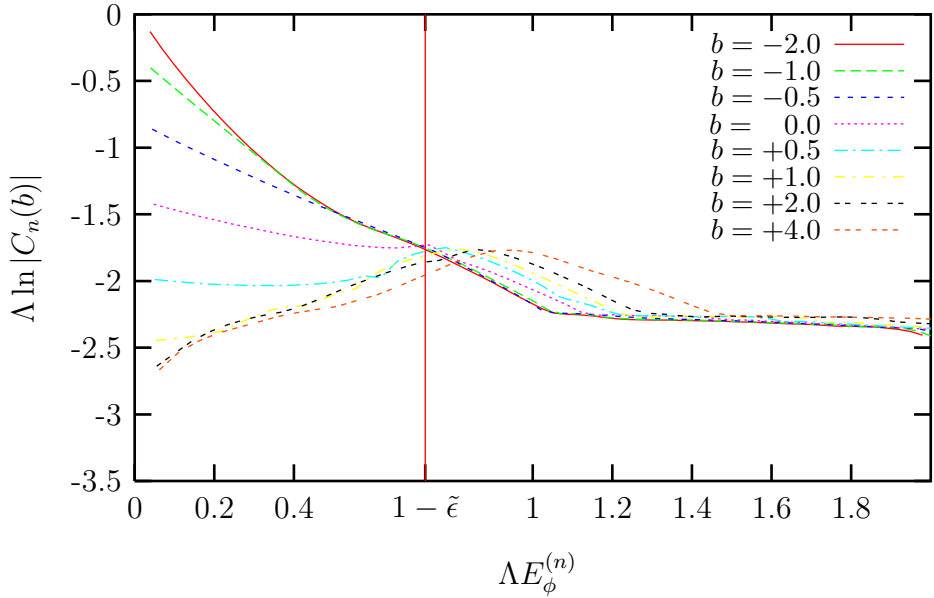 llx=0 lly=0 urx=720 ury=504 rwi=7200}
\put(3037,1247){\makebox(0,0)[r]{\small $b =  + 4.0$}}%
\put(3037,1347){\makebox(0,0)[r]{\small $b =  + 2.0$}}%
\put(3037,1447){\makebox(0,0)[r]{\small $b =  + 1.0$}}%
\put(3037,1547){\makebox(0,0)[r]{\small $b = +  0.5$}}%
\put(3037,1647){\makebox(0,0)[r]{\small $b = \;\;\;0.0$}}%
\put(3037,1747){\makebox(0,0)[r]{\small $b = - 0.5$}}%
\put(3037,1847){\makebox(0,0)[r]{\small $b = - 1.0$}}%
\put(3037,1947){\makebox(0,0)[r]{\small $b = - 2.0$}}%
\put(1950,-50){\makebox(0,0){ $\Lambda E_\phi^{(n)}$}}%
\put(100,1180){%
\special{ps: gsave currentpoint currentpoint translate
270 rotate neg exch neg exch translate}%
\makebox(0,0)[b]{\shortstack{$\Lambda \ln |C_n(b)|$}}%
\special{ps: currentpoint grestore moveto}%
}%
\put(3150,200){\makebox(0,0){$1.8$}}%
\put(2850,200){\makebox(0,0){$1.6$}}%
\put(2550,200){\makebox(0,0){1.4}}%
\put(2250,200){\makebox(0,0){$1.2$}}%
\put(1950,200){\makebox(0,0){1}}%
\put(1545,200){\makebox(0,0){$1 - \tilde\epsilon$}}%
\put(1050,200){\makebox(0,0){$0.4$}}%
\put(750,200){\makebox(0,0){$0.2$}}%
\put(450,200){\makebox(0,0){$0$}}%
\put(400,2060){\makebox(0,0)[r]{0}}%
\put(400,1809){\makebox(0,0)[r]{-0.5}}%
\put(400,1557){\makebox(0,0)[r]{-1}}%
\put(400,1306){\makebox(0,0)[r]{-1.5}}%
\put(400,1054){\makebox(0,0)[r]{-2}}%
\put(400,803){\makebox(0,0)[r]{-2.5}}%
\put(400,551){\makebox(0,0)[r]{-3}}%
\put(400,300){\makebox(0,0)[r]{-3.5}}%
\end{picture}%
\endgroup
 
\caption{\label{fig19} The same as in Fig.~\ref{fig17} but for
the interaction function $f_2$ 
and $\Lambda = 0.0256$, $M = 0.8$, $\omega = 0.6$, $\tilde{\epsilon} = 0.27$.}
\end{figure}

To describe our solutions at finite $b$, including classically
forbidden region, it is convenient to work in the basis of
the ``instantaneous'' Hamiltonian of the $\phi$-oscillator. The
eigenfunctions obey
\[
   \hat{{\cal H}}_\phi(b) \Psi_n^{(\Omega)} (b) = E_{\phi}^{(n)} (b) 
\Psi_n^{(\Omega)} (b)\;,
\]
where
\[
   E_\phi^{(n)} (b)= \Omega (b)\left( n+ \half \right)\;.
\]
The wave function is expanded 
\begin{equation}
  \Psi (b) = \sum_{n=0}^{\infty} C_n (b) \Psi_n^{(\Omega)} (b)\;.
\label{EqN1}
\end{equation}
We will be interested in the behavior of the occupation numbers
$|C_n(b)|^2$ as functions of the scale factor $b$ and excitation
number $n$. We will see that at small $\Lambda$, which is of primary
interest to us, the occupation numbers are exponential,
\[
 |C_n(b)|^2 \propto \mbox{e}^{\frac{2}{\Lambda} F(b, \Lambda E_n)}\;,
\]
so they are sharply peaked, at given $b$, near a certain
value of $n$. For sufficiently large values of the coupling parameter
$M$  and small $\Lambda$, typical plots of $\Lambda \ln |C_n|$
are shown in Figs.~\ref{fig17}  and \ref{fig19}. 

To discuss these plots, we recall first that the potential \eq{12**}
is centered at $b=0$ and its width equals $\sqrt{2}$, so the potential
is small at $b = \pm 2$, the first and last points displayed in
Fig.~\ref{fig17}. In Fig.~\ref{fig19} 
we show also the occupation numbers at
larger $b$ where the interaction function is large and additional
creation of $\phi$-quanta occurs, now in the classicaly allowed 
region.
We also note that we consider quite small  values of $\Lambda$, 
so the occupation numbers are
actually very small (because of tunneling suppression) in the
forbidden region and behind the barrier. Nevertheless we are confident
that the results are reliable, partially due to checks of our
numerical procedure summarized in Appendix B. 

It is clear from Figs.~\ref{fig17} and \ref{fig19} that on the left of the
barrier, the $\phi$-oscillator is mostly in its ground state (since we
have chosen $n_0 =0$), while in the classically forbidden region its
wave function gradually becomes peaked at the value of $n$ such that
the total matter energy, $(E_\phi^{(n)} + \tilde{\epsilon}/\Lambda)$,
is close to the height of the barrier, $V_{max} =
1/\Lambda$. Comparing Figs.~\ref{fig17} and \ref{fig19}, 
we see that this property
is essentially independent of the shape of the interaction function,
as well as on parameters of the model (provided that $M$ is
sufficiently large, see below).
We have checked that this is indeed the case by studying the solutions
for numerous sets of parameters and also 
for the interaction function $f_3$. 

The mechanism due to which tunneling is accompanied by
the creation of $\phi$-quanta is clear. The $\phi$-$b$ interaction 
results in weak creation of these quanta (either in front of the 
barrier, as is the case for the interaction function $f_1$, or
in the forbidden region, as is the case for $f_2$ and $f_3$).
The low modes get strongly suppressed in the forbidden region,
while the modes of energy $E^{(n)}_\phi \sim 1/\Lambda$
survive. The latter effect occurs because the total energy
of matter is close to the barrier height. Hence, there is strong
back reaction of created $\phi$-quanta on tunneling.

One remark is in order. Even though we impose the boundary condition
\eq{24*} with $n_0=0$, the wave function on the left of the barrier
still contains a small admixture of the excited states of the
$\phi$-oscillator. This is due to the  presence of the reflected wave.
 
Another way of studying the properties of our solutions is to consider
the average energy of the $\phi$-oscillator. For finite $b$ we define
it as follows,
\be
\langle E_\phi \rangle (b) = 
\frac{\sum\limits_{n=0}^{\infty} 
E_\phi^{(n)} (b) |C_n (b)|^2}{\sum\limits_{n=0}^{\infty}\;. 
|C_n (b)|^2}
\label{28*}
\ee
In the asymptotic region $b \to +\infty$, a more appropriate
definition would involve the flux factor, as in conventional
quantum mechanics,
\be
\langle E_\phi \rangle^{asymp} = 
\left.\frac{\sum\limits_{n=0}^{\infty} 
k_n \omega(n+ 1/2) |C_n|^2}{\sum\limits_{n=0}^{\infty} 
k_n |C_n|^2} \right|_{b \to + \infty}\;.
\label{28**}
\ee
Since $|C_n|$ are sharply peaked at a certain $n$, the two definitions
do not differ much: 
$\langle E_\phi \rangle^{asymp} \approx 
\langle E_\phi \rangle (b \to +\infty)$. We will use the definition
\eq{28*} for general $b$, and the definition \eq{28**} when discussing
the asymptotics $b \to +\infty$.
\begin{figure}[ht!]
\begingroup%
  \makeatletter%
  \newcommand{\GNUPLOTspecial}{%
    \@sanitize\catcode`\%=14\relax\special}%
  \setlength{\unitlength}{0.1bp}%
\begin{picture}(3600,2160)(0,0)%
\special{psfile=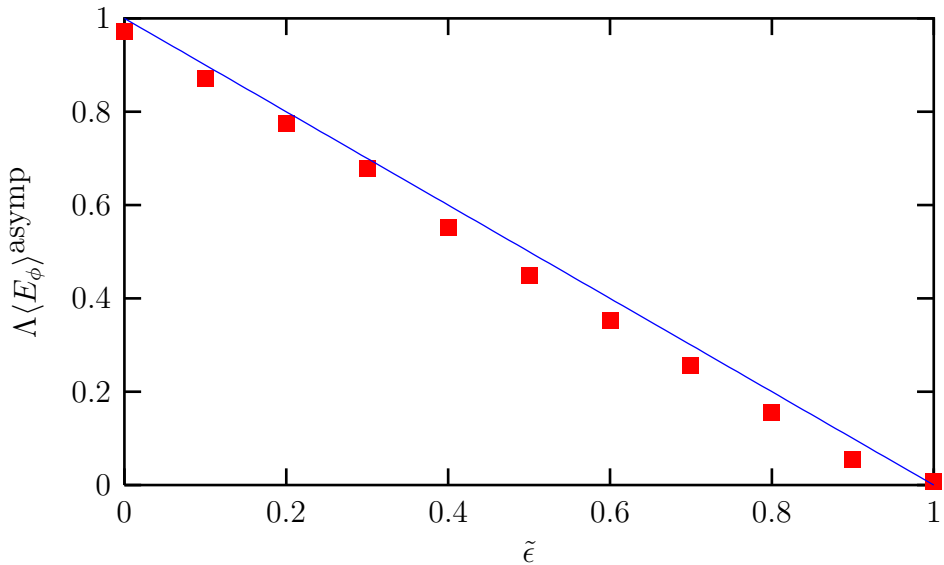 llx=0 lly=0 urx=720 ury=504 rwi=7200}
\put(1925,50){\makebox(0,0){$\tilde\epsilon$}}%
\put(100,1180){%
\special{ps: gsave currentpoint currentpoint translate
270 rotate neg exch neg exch translate}%
\makebox(0,0)[b]{\shortstack{$ \Lambda \langle E_\phi\rangle^{\mbox{asymp}}$}}%
\special{ps: currentpoint grestore moveto}%
}%
\put(3450,200){\makebox(0,0){1}}%
\put(2840,200){\makebox(0,0){0.8}}%
\put(2230,200){\makebox(0,0){0.6}}%
\put(1620,200){\makebox(0,0){0.4}}%
\put(1010,200){\makebox(0,0){0.2}}%
\put(400,200){\makebox(0,0){0}}%
\put(350,2060){\makebox(0,0)[r]{1}}%
\put(350,1708){\makebox(0,0)[r]{0.8}}%
\put(350,1356){\makebox(0,0)[r]{0.6}}%
\put(350,1004){\makebox(0,0)[r]{0.4}}%
\put(350,652){\makebox(0,0)[r]{0.2}}%
\put(350,300){\makebox(0,0)[r]{0}}%
\end{picture}%
\endgroup
 
\caption{\label{fig6} $\phi$-oscillator energy at large $b$ as
function of $\tilde{\epsilon}$ for the interaction function $f_1$ and
$\Lambda = 0.0256$, $M = 0.2$, $\omega = 0.6$.}
\end{figure}

\begin{figure}[ht!]
\begingroup%
  \makeatletter%
  \newcommand{\GNUPLOTspecial}{%
    \@sanitize\catcode`\%=14\relax\special}%
  \setlength{\unitlength}{0.1bp}%
\begin{picture}(3600,2160)(0,0)%
\special{psfile=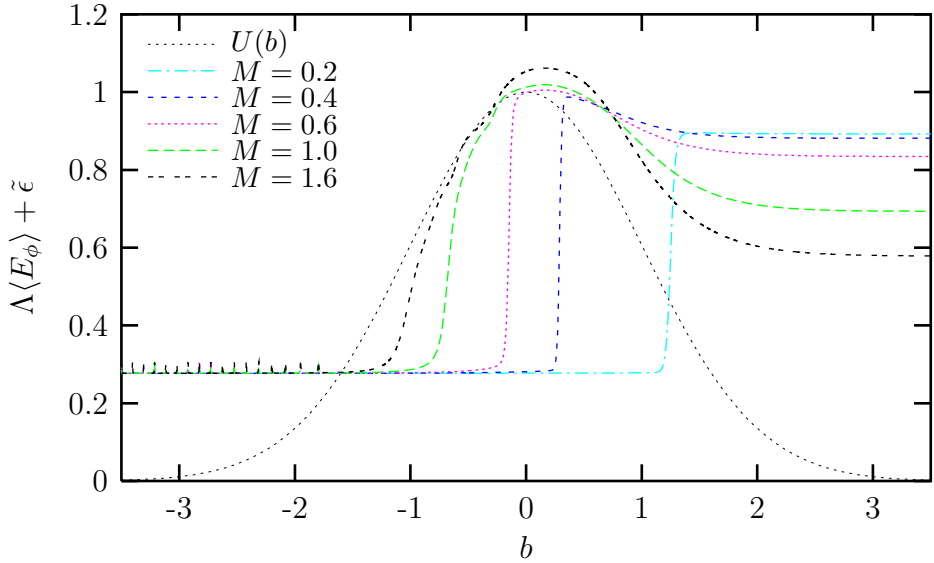 llx=0 lly=0 urx=720 ury=504 rwi=7200}
\put(813,1447){\makebox(0,0)[l]{\small $M = 1.6$}}%
\put(813,1547){\makebox(0,0)[l]{\small $M = 1.0$}}%
\put(813,1647){\makebox(0,0)[l]{\small $M = 0.6$}}%
\put(813,1747){\makebox(0,0)[l]{\small $M = 0.4$}}%
\put(813,1847){\makebox(0,0)[l]{\small $M = 0.2$}}%
\put(813,1947){\makebox(0,0)[l]{\small $ U(b)$}}%
\put(1925,50){\makebox(0,0){$b$}}%
\put(100,1180){%
\special{ps: gsave currentpoint currentpoint translate
270 rotate neg exch neg exch translate}%
\makebox(0,0)[b]{\shortstack{$\Lambda \langle E_\phi\rangle + \tilde\epsilon$}}%
\special{ps: currentpoint grestore moveto}%
}%
\put(3232,200){\makebox(0,0){3}}%
\put(2796,200){\makebox(0,0){2}}%
\put(2361,200){\makebox(0,0){1}}%
\put(1925,200){\makebox(0,0){0}}%
\put(1489,200){\makebox(0,0){-1}}%
\put(1054,200){\makebox(0,0){-2}}%
\put(618,200){\makebox(0,0){-3}}%
\put(350,2060){\makebox(0,0)[r]{1.2}}%
\put(350,1767){\makebox(0,0)[r]{1}}%
\put(350,1473){\makebox(0,0)[r]{0.8}}%
\put(350,1180){\makebox(0,0)[r]{0.6}}%
\put(350,887){\makebox(0,0)[r]{0.4}}%
\put(350,593){\makebox(0,0)[r]{0.2}}%
\put(350,300){\makebox(0,0)[r]{0}}%
\end{picture}%
\endgroup
 
\caption{\label{fig8} Total matter energy in units of the barrier
height, 
as function of
$b$ for different $M$; the interaction function is $f_3$ and
$\Lambda = 0.0225 $, $\omega= 0.6$, $\tilde{\epsilon} = 0.27$.}
\end{figure}

As one anticipates from Figs.~\ref{fig17} and \ref{fig19}, 
the total matter
energy at $b \to +\infty$ should be close to $V_{max} = 1/\Lambda$,
i.e., 
\be
  \langle E_\phi \rangle^{asymp} \approx \frac{1}{\Lambda}
(1-\tilde{\epsilon})\;.
\label{29*}
\ee
This is shown in Fig.~\ref{fig6}. 
In fact, the relation \eq{29*} should not
be precise, because the relevant quantity is the oscillator
energy near the top of the barrier rather than its asymptotic
value. For given $n$ the former is equal to $(n+1/2) \Omega (b\approx
0) = (n+1/2)\sqrt{\omega^2 + M^2 f(b\approx 0)}$ which is larger than
the asymptotic value $(n+1/2)\omega$. Hence, it is more appropriate
to study the oscillator energy \eq{28*} as function of $b$ for
different values of parameters. The corresponding plots are shown in
Fig.~\ref{fig8}, which clearly displays how the system climbs up the
potential barrier, and then stays at constant $|C_n|^2$, with energy
decreasing due to the  effect just described.

Figure~\ref{fig8} 
reveals another property: at smaller interaction strength
$M$, the $\phi$-oscillator gets excited ``later'', i.e., at larger
$b$. This may be anticipated from eq.~\eq{16+}, as the exponential factor
compensates for small $M$ at larger $b$. In fact, the properties we
discuss should be absent for very small $M$, and the perturbative
regime of section 2.2 should set in instead. We will come back to this
point later on.

\begin{figure}[ht!]
\begingroup%
  \makeatletter%
  \newcommand{\GNUPLOTspecial}{%
    \@sanitize\catcode`\%=14\relax\special}%
  \setlength{\unitlength}{0.1bp}%
\begin{picture}(3600,2160)(0,0)%
\special{psfile=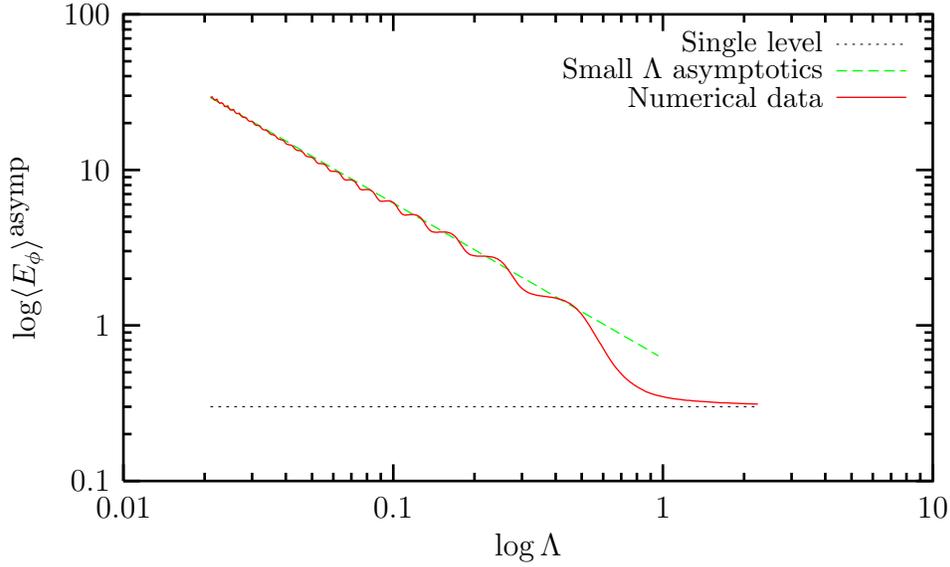 llx=0 lly=0 urx=720 ury=504 rwi=7200}
\put(3037,1747){\makebox(0,0)[r]{\small Numerical data}}%
\put(3037,1847){\makebox(0,0)[r]{\small Small $\Lambda$ asymptotics}}%
\put(3037,1947){\makebox(0,0)[r]{\small Single level}}%
\put(1925,50){\makebox(0,0){$\log \Lambda$}}%
\put(100,1180){%
\special{ps: gsave currentpoint currentpoint translate
270 rotate neg exch neg exch translate}%
\makebox(0,0)[b]{\shortstack{$\log \langle E_\phi\rangle^{\mbox{\small asymp}}$}}%
\special{ps: currentpoint grestore moveto}%
}%
\put(3450,200){\makebox(0,0){10}}%
\put(2433,200){\makebox(0,0){1}}%
\put(1417,200){\makebox(0,0){0.1}}%
\put(400,200){\makebox(0,0){0.01}}%
\put(350,2060){\makebox(0,0)[r]{100}}%
\put(350,1473){\makebox(0,0)[r]{10}}%
\put(350,887){\makebox(0,0)[r]{1}}%
\put(350,300){\makebox(0,0)[r]{0.1}}%
\end{picture}%
\endgroup
 
\caption{\label{fig1} $\phi$-oscillator energy at $b \to +\infty$
as function of $\Lambda$, in log-log scale. Straight lines are
asymptotics \eq{30+} and \eq{30*}. The interaction function is
$f_3$, the parameters are:
$M=0.4$, $\omega=0.6$, $\tilde{\epsilon} =0.27$.}
\end{figure}

\begin{figure}[ht!]
\begingroup%
  \makeatletter%
  \newcommand{\GNUPLOTspecial}{%
    \@sanitize\catcode`\%=14\relax\special}%
  \setlength{\unitlength}{0.1bp}%
\begin{picture}(3600,2160)(0,0)%
\special{psfile=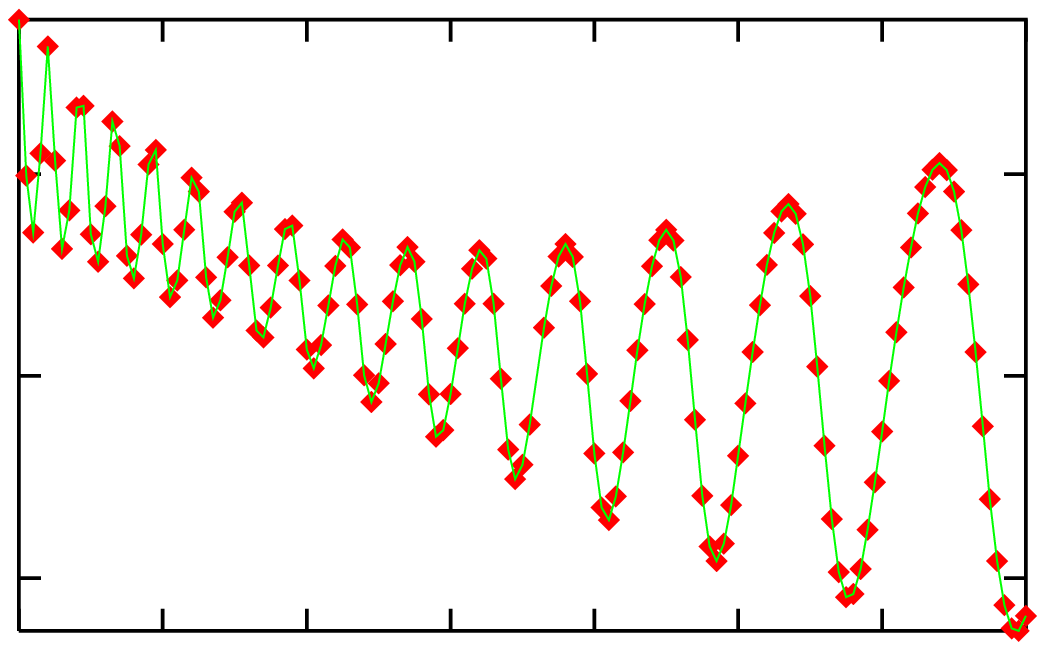 llx=0 lly=0 urx=720 ury=504 rwi=7200}
\put(2000,50){\makebox(0,0){$\sqrt\Lambda$}}%
\put(100,1180){%
\special{ps: gsave currentpoint currentpoint translate
270 rotate neg exch neg exch translate}%
\makebox(0,0)[b]{\shortstack{$\Lambda \langle E_\phi\rangle^{\mbox{\small asymp}}$}}%
\special{ps: currentpoint grestore moveto}%
}%
\put(3450,200){\makebox(0,0){$0.30$}}%
\put(3036,200){\makebox(0,0){$0.28$}}%
\put(2621,200){\makebox(0,0){$0.26$}}%
\put(2207,200){\makebox(0,0){$0.24$}}%
\put(1793,200){\makebox(0,0){$0.22$}}%
\put(1379,200){\makebox(0,0){$0.20$}}%
\put(964,200){\makebox(0,0){$0.18$}}%
\put(550,200){\makebox(0,0){$0.16$}}%
\put(500,1615){\makebox(0,0)[r]{$0.61$}}%
\put(500,1034){\makebox(0,0)[r]{$0.59$}}%
\put(500,452){\makebox(0,0)[r]{$0.57$}}%
\end{picture}%
\endgroup
 
\caption{\label{fig2} The same as in Fig.~\ref{fig1}, but for a region of
small $\Lambda$. The scales are linear.}
\end{figure}

It is instructive to see what happens at different $\Lambda$.
We still consider tunneling from the oscillator ground state,
$n_0 =0$. At large $\Lambda$ the barrier is effectively absent, so the
$\phi$-oscillator remains in its ground state with energy
\be
      E_\phi^{(0)} = \half \Omega\;.
\label{30+}
\ee
At small $\Lambda$ the energy of the $\phi$-oscillator increases
substantially, to
\be
       \langle E_\phi \rangle^{asymp} \approx 
\frac{\mbox{const}}{\Lambda}\;.
\label{30*}
\ee
The overall behavior of the asymptotic value of the oscillator energy
is shown as function of $\Lambda$ in Fig.~\ref{fig1} 
in log-log scale. One
indeed observes the asymptotics \eq{30*} at small $\Lambda$
and \eq{30+} at large $\Lambda$. 
The oscillatory behavior of $\langle E_\phi \rangle^{asymp}$ as
function of $\Lambda$, shown in detail in Fig.~\ref{fig2}, 
may be understood
as follows. For finite $\Lambda$, the number of ``under-barrier''
oscillator levels (which have  
energy below $(1-\tilde{\epsilon})/\Lambda$) is finite, and as
$\Lambda$ decreases, some over-barrier levels become under-barrier.
Since the system tends to climb on the top of the barrier, but not
much above the top, the
structure of levels near the top matters. This structure changes with 
$\Lambda$ in an oscillatory way, hence the oscillatory behavior shown
in Fig.~\ref{fig2}.

\begin{figure}[ht!]
\begingroup%
  \makeatletter%
  \newcommand{\GNUPLOTspecial}{%
    \@sanitize\catcode`\%=14\relax\special}%
  \setlength{\unitlength}{0.1bp}%
\begin{picture}(3600,2160)(0,0)%
\special{psfile=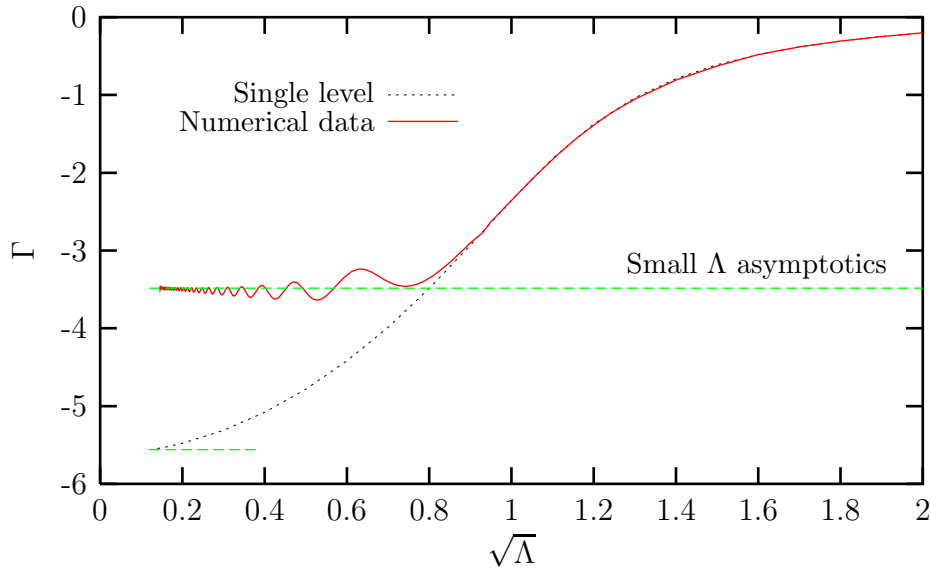 llx=0 lly=0 urx=720 ury=504 rwi=7200}
\put(1385,1667){\makebox(0,0)[r]{\small Numerical data}}%
\put(1385,1767){\makebox(0,0)[r]{\small Single level}}%
\put(2334,1121){\makebox(0,0)[l]{\small Small $\Lambda$ asymptotics}}%
\put(1900,50){\makebox(0,0){$\sqrt\Lambda$}}%
\put(100,1180){%
\special{ps: gsave currentpoint currentpoint translate
270 rotate neg exch neg exch translate}%
\makebox(0,0)[b]{\shortstack{$\Gamma$}}%
\special{ps: currentpoint grestore moveto}%
}%
\put(3450,200){\makebox(0,0){2}}%
\put(3140,200){\makebox(0,0){1.8}}%
\put(2830,200){\makebox(0,0){1.6}}%
\put(2520,200){\makebox(0,0){1.4}}%
\put(2210,200){\makebox(0,0){1.2}}%
\put(1900,200){\makebox(0,0){1}}%
\put(1590,200){\makebox(0,0){0.8}}%
\put(1280,200){\makebox(0,0){0.6}}%
\put(970,200){\makebox(0,0){0.4}}%
\put(660,200){\makebox(0,0){0.2}}%
\put(350,200){\makebox(0,0){0}}%
\put(300,2060){\makebox(0,0)[r]{0}}%
\put(300,1767){\makebox(0,0)[r]{-1}}%
\put(300,1473){\makebox(0,0)[r]{-2}}%
\put(300,1180){\makebox(0,0)[r]{-3}}%
\put(300,887){\makebox(0,0)[r]{-4}}%
\put(300,593){\makebox(0,0)[r]{-5}}%
\put(300,300){\makebox(0,0)[r]{-6}}%
\end{picture}%
\endgroup
 
\caption{\label{fig3} The magnitude of the wave function behind the
barrier as function of $\sqrt{\Lambda}$. The quantity $\Gamma$ is
defined in eq.~\eq{31a+}. Short dashed line in the lower left part
corresponds to the naive exponent \eq{31a**}. The interaction function 
and parameters are the same as in Fig.~\ref{fig8}.}
\end{figure}

A similar dependence on $\Lambda$ is inherent in the magnitude of the
wave function behind the barrier. Let us define a quantity $\Gamma$ as
follows\footnote{In conventional quantum mechanics, 
$\exp (\Gamma / \Lambda)$ would be 
tunneling probability, while in quantum cosmology such an
interpretation may be debatable.},
\be
\mbox{e}^{\frac{\Gamma}{\Lambda}} = \left.\sum_{n=0}^{\infty}
|C_n|^2 \right|_{b \to +\infty}\;,
\label{31a+}
\ee
where we again consider tunneling from the ground state of the
$\phi$-oscillator. 
We find numerically
that at small $\Lambda$ the
exponent scales as 
\be
     \Gamma = \Gamma (M, \omega)\;,
\label{31a*}
\ee
i.e., it is independent of $\Lambda$. This is shown in Fig.~\ref{fig3}.
Note that in the regime we consider, $\Gamma / \Lambda$ 
is much larger than the
naive tunneling exponent 
\be
\mbox{e}^{ - \displaystyle\int_{a_1}^{a_2}~da~\sqrt{2 (V(a)- \epsilon)}}\;,
\label{31a**}
\ee
 entering
eq.~\eq{4*}. Thus, creation of $\phi$-quanta has dramatic effect on
the magnitude of the wave function behind the barrier.

\begin{figure}[ht!]
\begingroup%
  \makeatletter%
  \newcommand{\GNUPLOTspecial}{%
    \@sanitize\catcode`\%=14\relax\special}%
  \setlength{\unitlength}{0.1bp}%
\begin{picture}(3600,2160)(0,0)%
\special{psfile=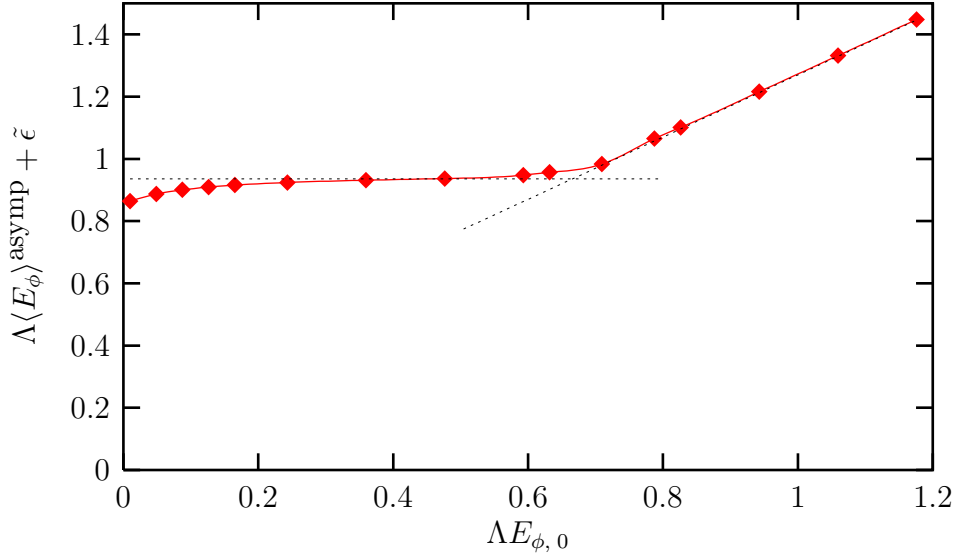 llx=0 lly=0 urx=720 ury=504 rwi=7200}
\put(1925,50){\makebox(0,0){$\Lambda E_{\phi,\;0} $}}%
\put(100,1180){%
\special{ps: gsave currentpoint currentpoint translate
270 rotate neg exch neg exch translate}%
\makebox(0,0)[b]{\shortstack{$ \Lambda \langle E_\phi\rangle^{\mbox{\small asymp}} + \tilde\epsilon$}}%
\special{ps: currentpoint grestore moveto}%
}%
\put(3450,200){\makebox(0,0){1.2}}%
\put(2942,200){\makebox(0,0){1}}%
\put(2433,200){\makebox(0,0){0.8}}%
\put(1925,200){\makebox(0,0){0.6}}%
\put(1417,200){\makebox(0,0){0.4}}%
\put(908,200){\makebox(0,0){0.2}}%
\put(400,200){\makebox(0,0){0}}%
\put(350,1943){\makebox(0,0)[r]{1.4}}%
\put(350,1708){\makebox(0,0)[r]{1.2}}%
\put(350,1473){\makebox(0,0)[r]{1}}%
\put(350,1239){\makebox(0,0)[r]{0.8}}%
\put(350,1004){\makebox(0,0)[r]{0.6}}%
\put(350,769){\makebox(0,0)[r]{0.4}}%
\put(350,535){\makebox(0,0)[r]{0.2}}%
\put(350,300){\makebox(0,0)[r]{0}}%
\end{picture}%
\endgroup
 
\caption{\label{fig16} $\phi$-oscillator energy at $b \to +\infty$
as function of the oscillator energy on the left of the barrier for
interaction function $f_3$ and $\Lambda = 0.0324$, $M = 0.2$, 
$\tilde{\epsilon} = 0.27$. 
The linear behavior at large $E_{\phi,0}$ corresponds to classically
allowed transitions.}
\end{figure}

In the regime of relatively large interaction parameter $M$, the
dominant part of the wave function behind the barrier corresponds to
$\phi$-oscillator excited to a particular energy, such that the total
matter energy near the top of the barrier is close to $V_{max} =
1/\Lambda$. Hence, this state is essentially independent of the
state of the $\phi$-oscillator on the left of the
barrier, i.e., of $n_0$, up to the overall
penetration factor (the latter depends on $n_0$ strongly). This is
illustrated in Fig.~\ref{fig16}.

\begin{figure}[ht!]
\begingroup%
  \makeatletter%
  \newcommand{\GNUPLOTspecial}{%
    \@sanitize\catcode`\%=14\relax\special}%
  \setlength{\unitlength}{0.1bp}%
\begin{picture}(3600,2160)(0,0)%
\special{psfile=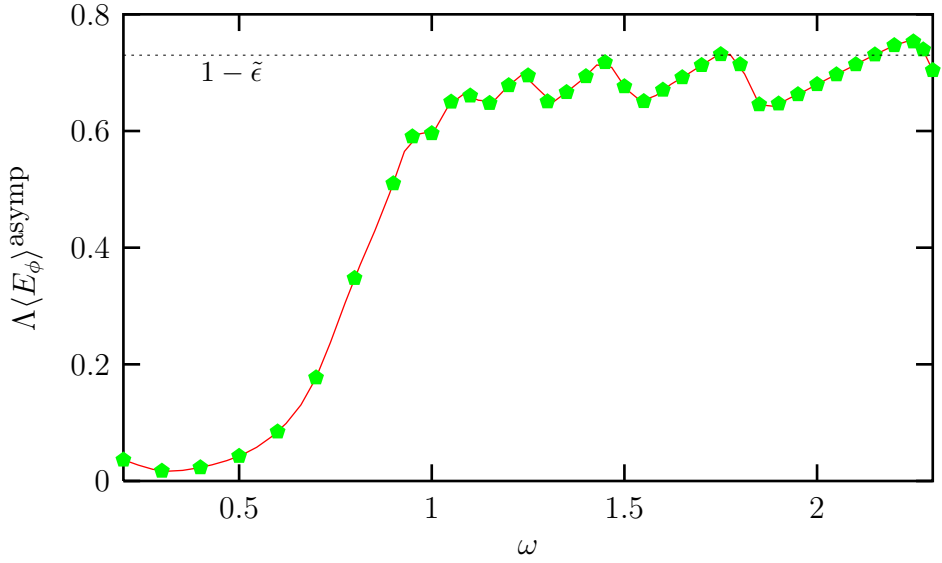 llx=0 lly=0 urx=720 ury=504 rwi=7200}
\put(690,1840){\makebox(0,0)[l]{\small $1 - \tilde\epsilon$}}%
\put(1925,50){\makebox(0,0){$\omega$}}%
\put(100,1180){%
\special{ps: gsave currentpoint currentpoint translate
270 rotate neg exch neg exch translate}%
\makebox(0,0)[b]{\shortstack{$\Lambda \langle E_\phi\rangle^{\mbox{\small asymp}}$}}%
\special{ps: currentpoint grestore moveto}%
}%
\put(3014,200){\makebox(0,0){2}}%
\put(2288,200){\makebox(0,0){1.5}}%
\put(1562,200){\makebox(0,0){1}}%
\put(836,200){\makebox(0,0){0.5}}%
\put(350,2060){\makebox(0,0)[r]{0.8}}%
\put(350,1620){\makebox(0,0)[r]{0.6}}%
\put(350,1180){\makebox(0,0)[r]{0.4}}%
\put(350,740){\makebox(0,0)[r]{0.2}}%
\put(350,300){\makebox(0,0)[r]{0}}%
\end{picture}%
\endgroup
 
\caption{\label{fig20} $\phi$-oscillator energy at $b \to +\infty$ as
function of $\omega$ for the interaction function $f_2$, $M = 0.3$,
$\Lambda = 0.04$, $\tilde\epsilon = 0.27$  
and $n_0= 0$. At small $\omega$ the oscillator does not
get excited, while at larger $\omega$ it is strongly excited due to
tunneling. Oscillations at large $\omega$ have the same origin as
oscillations in Figs.~\ref{fig1}, \ref{fig2} and \ref{fig3}.}
\end{figure}

Let us now discuss the case of small $M$, when the couping between
$\phi$ and $b$ is weak. According to the perturbative  analysis
of section 2.2 in that case the $\phi$-oscillator does not get
excited due to tunneling, provided that $\omega$ is small. However,
one may expect that at
large enough $\omega$ the situation is different and that the system is
back in the regime of strong particle creation. 
Hence, as $\omega$ increases,
the asymptotic energy of the oscillator should change from a small
value to 
$\langle E_{\phi} \rangle^{asymp} \approx (1-\tilde{\epsilon})/\Lambda$. 
This is shown in Fig.~\ref{fig20}. 
Of course, the value of $\omega$ at which
the transition between the two regimes occurs depends on $M$ and other
parameters, but in any case the scalar quanta with large enough
$\omega$ are copiously created due to tunneling. In the context of
quantum cosmology this means that the Universe after tunneling is
filled with particles, possibly of large mass and high
spatial momenta.

\subsection{Scaling with $\Lambda$.}

We have seen in section 3.2 that at small $\Lambda$ and sufficiently
large $M$ and/or $\omega$, the $\phi$-oscillator gets excited in such
a way that the total energy of matter is close to the height of the
barrier. This means that the wave function of the oscillator is peaked
near a certain excitation number $n \propto 1/ \Lambda$. One may
consider this property as an indication that the tunneling transition
is of semiclassical type. Further support of this conjecture comes
from the semiclassical-type scaling of the overall magnitude of the
wave function behind the barrier, eq.~\eq{31a*}. Let us describe
further empirical evidence in favor of the semiclassical nature of the
tunneling transitions in the regime where the $\phi$-oscillator gets
strongly excited.

\begin{figure}[ht!]
\begingroup%
  \makeatletter%
  \newcommand{\GNUPLOTspecial}{%
    \@sanitize\catcode`\%=14\relax\special}%
  \setlength{\unitlength}{0.1bp}%
\begin{picture}(3600,2160)(0,0)%
\special{psfile=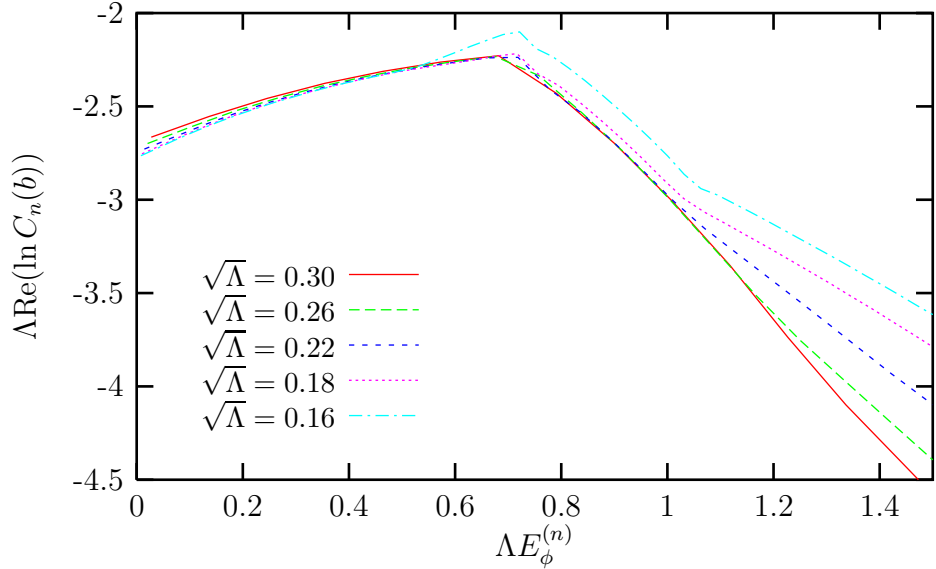 llx=0 lly=0 urx=720 ury=504 rwi=7200}
\put(1200,542){\makebox(0,0)[r]{\small $\sqrt\Lambda  = 0.16$}}%
\put(1200,675){\makebox(0,0)[r]{\small $\sqrt\Lambda  = 0.18$}}%
\put(1200,808){\makebox(0,0)[r]{\small $\sqrt\Lambda  = 0.22$}}%
\put(1200,941){\makebox(0,0)[r]{\small $\sqrt\Lambda  = 0.26$}}%
\put(1200,1074){\makebox(0,0)[r]{\small $\sqrt\Lambda = 0.30$}}%
\put(1950,50){\makebox(0,0){$\Lambda E_\phi^{(n)}$}}%
\put(100,1180){%
\special{ps: gsave currentpoint currentpoint translate
270 rotate neg exch neg exch translate}%
\makebox(0,0)[b]{\shortstack{$\Lambda \mbox{Re}(\ln C_n(b))$}}%
\special{ps: currentpoint grestore moveto}%
}%
\put(3250,200){\makebox(0,0){1.4}}%
\put(2850,200){\makebox(0,0){1.2}}%
\put(2450,200){\makebox(0,0){1}}%
\put(2050,200){\makebox(0,0){0.8}}%
\put(1650,200){\makebox(0,0){0.6}}%
\put(1250,200){\makebox(0,0){0.4}}%
\put(850,200){\makebox(0,0){0.2}}%
\put(450,200){\makebox(0,0){0}}%
\put(400,2060){\makebox(0,0)[r]{-2}}%
\put(400,1708){\makebox(0,0)[r]{-2.5}}%
\put(400,1356){\makebox(0,0)[r]{-3}}%
\put(400,1004){\makebox(0,0)[r]{-3.5}}%
\put(400,652){\makebox(0,0)[r]{-4}}%
\put(400,300){\makebox(0,0)[r]{-4.5}}%
\end{picture}%
\endgroup
 
\caption{\label{fig14} Real part of the exponent of the oscillator wave
function behind the barrier at
different
$\Lambda$ for interaction function $f_1$ and 
$M=0.2$, $\omega = 0.6$, $\tilde{\epsilon} = 0.27$.
Note that the wave functions themselves differ by more than 10 orders
of magnitude at different values of $\Lambda$.}
\end{figure}

\begin{figure}[ht!]
\begingroup%
  \makeatletter%
  \newcommand{\GNUPLOTspecial}{%
    \@sanitize\catcode`\%=14\relax\special}%
  \setlength{\unitlength}{0.1bp}%
\begin{picture}(3600,2160)(0,0)%
\special{psfile=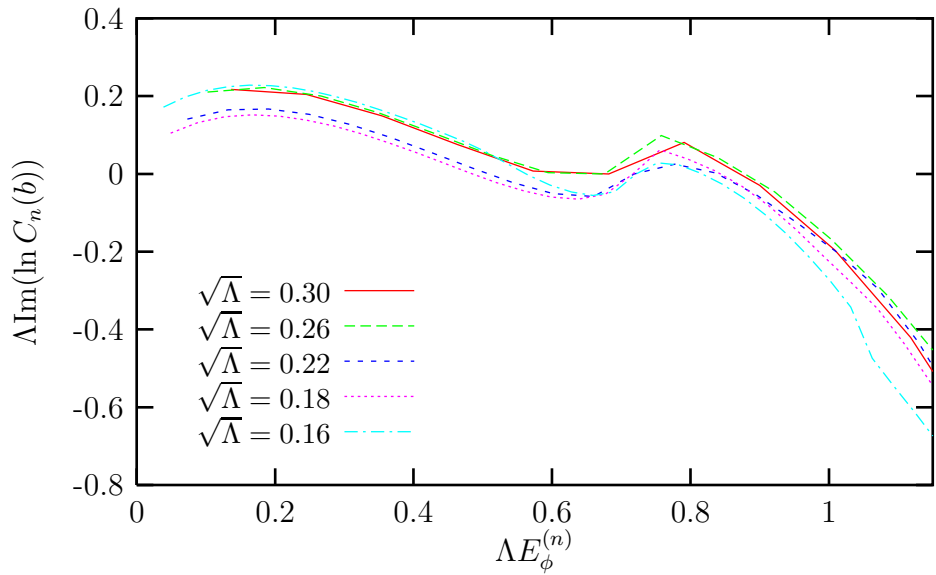 llx=0 lly=0 urx=720 ury=504 rwi=7200}
\put(1183,501){\makebox(0,0)[r]{\small $\sqrt\Lambda = 0.16$}}%
\put(1183,634){\makebox(0,0)[r]{\small $\sqrt\Lambda = 0.18$}}%
\put(1183,767){\makebox(0,0)[r]{\small $\sqrt\Lambda = 0.22$}}%
\put(1183,900){\makebox(0,0)[r]{\small $\sqrt\Lambda = 0.26$}}%
\put(1183,1033){\makebox(0,0)[r]{\small $\sqrt\Lambda = 0.30$}}%
\put(1950,50){\makebox(0,0){$\Lambda E_\phi^{(n)}$}}%
\put(100,1180){%
\special{ps: gsave currentpoint currentpoint translate
270 rotate neg exch neg exch translate}%
\makebox(0,0)[b]{\shortstack{$\Lambda \hbox{Im} (\ln C_n(b))$}}%
\special{ps: currentpoint grestore moveto}%
}%
\put(3059,200){\makebox(0,0){1}}%
\put(2537,200){\makebox(0,0){0.8}}%
\put(2015,200){\makebox(0,0){0.6}}%
\put(1493,200){\makebox(0,0){0.4}}%
\put(972,200){\makebox(0,0){0.2}}%
\put(450,200){\makebox(0,0){0}}%
\put(400,2060){\makebox(0,0)[r]{0.4}}%
\put(400,1767){\makebox(0,0)[r]{0.2}}%
\put(400,1473){\makebox(0,0)[r]{0}}%
\put(400,1180){\makebox(0,0)[r]{-0.2}}%
\put(400,887){\makebox(0,0)[r]{-0.4}}%
\put(400,593){\makebox(0,0)[r]{-0.6}}%
\put(400,300){\makebox(0,0)[r]{-0.8}}%
\end{picture}%
\endgroup
 
\caption{\label{fig15} The same  as in Fig.~\ref{fig14}, but for imaginary
part.}
\end{figure}

One observation is that not only the overall magnitude, but also the
entire wave function has the semiclassical form 
\be
   C_n (b) 
= \mbox{e}^{\frac{1}{\Lambda} {\cal F}(b, \Lambda E_\phi^{(n)})}\;,
\label{35*}
\ee
both in the classically forbidden region and behind the barrier. As an
example, we show in Figs.~\ref{fig14} and \ref{fig15} 
real and imaginary parts of
${\cal F} \equiv \Lambda \ln C_n$ behind the barrier
as functions of 
$\Lambda E_\phi^{(n)}$ 
at different values of $\Lambda$. Scaling property
\eq{35*} is clearly seen in these figures. We observed very similar 
scaling  of the wave function 
in the classically forbidden region, though 
its shape is quite different there
(the latter is shown for a representative 
value of $\Lambda$ in Figs.~\ref{fig17} and \ref{fig19}). 


\begin{figure}[ht!]
\begingroup%
  \makeatletter%
  \newcommand{\GNUPLOTspecial}{%
    \@sanitize\catcode`\%=14\relax\special}%
  \setlength{\unitlength}{0.1bp}%
\begin{picture}(3600,2160)(0,0)%
\special{psfile=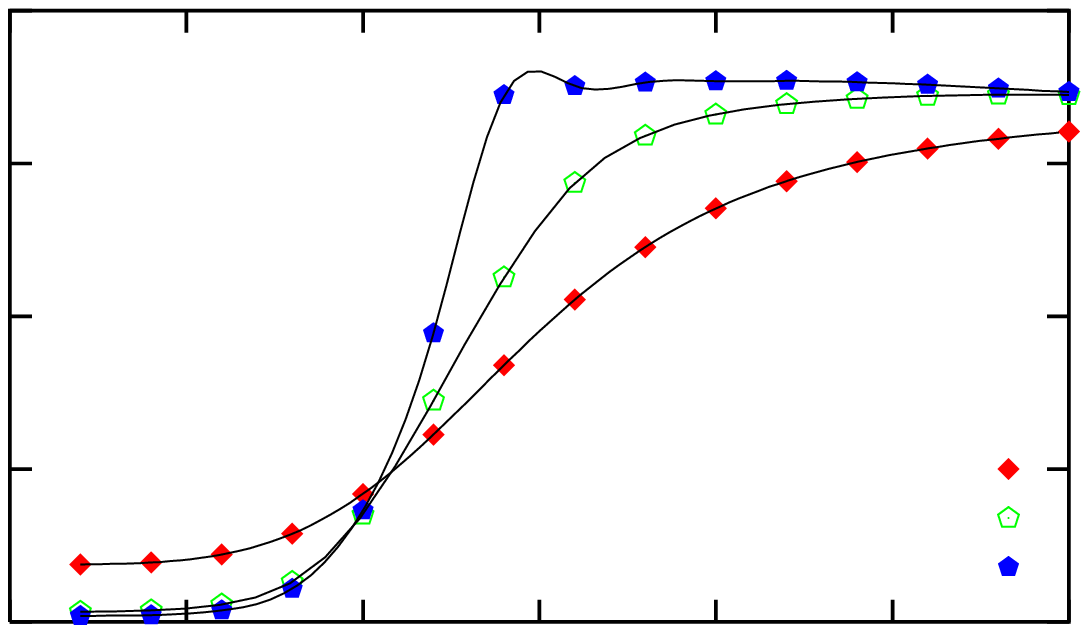 llx=0 lly=0 urx=720 ury=504 rwi=7200}
\put(3095,458){\makebox(0,0)[r]{\small $\sqrt\Lambda = 0.16$}}%
\put(3095,599){\makebox(0,0)[r]{\small $\sqrt\Lambda = 0.21$}}%
\put(3095,740){\makebox(0,0)[r]{\small $\sqrt\Lambda = 0.50$}}%
\put(1925,50){\makebox(0,0){$M$}}%
\put(100,1180){%
\special{ps: gsave currentpoint currentpoint translate
270 rotate neg exch neg exch translate}%
\makebox(0,0)[b]{\shortstack{$\Lambda \langle E_\phi\rangle^{\mbox{\small asymp}}$}}%
\special{ps: currentpoint grestore moveto}%
}%
\put(3450,200){\makebox(0,0){0.3}}%
\put(2942,200){\makebox(0,0){0.25}}%
\put(2433,200){\makebox(0,0){0.2}}%
\put(1925,200){\makebox(0,0){0.15}}%
\put(1417,200){\makebox(0,0){0.1}}%
\put(908,200){\makebox(0,0){0.05}}%
\put(400,200){\makebox(0,0){0}}%
\put(350,2060){\makebox(0,0)[r]{0.8}}%
\put(350,1620){\makebox(0,0)[r]{0.6}}%
\put(350,1180){\makebox(0,0)[r]{0.4}}%
\put(350,740){\makebox(0,0)[r]{0.2}}%
\put(350,300){\makebox(0,0)[r]{0}}%
\end{picture}%
\endgroup
 
\caption{\label{fig7}
$\phi$-oscillator energy behind the barrier as function of $M$ at
different $\Lambda$ for the interaction function $f_1$ and
$\omega =0.6$, $\tilde{\epsilon} = 0.27$.}
\end{figure}

More evidence comes from the study of the transition from the
perturbative regime, occuring at small values of the coupling
parameter $M$, to the regime of strong excitation of the
$\phi$-oscillator at larger $M$. 
As shown in Fig.~\ref{fig7}, this transition becomes more
sharp as $\Lambda$ decreases, and becomes essentially step function at
small $\Lambda$. We consider this as an indication that the wave
function behind the barrier and in the forbidden region is a
combination of two exponentials of semiclassical type, eq.~\eq{10*}.

What classical solutions describe  the tunneling
transitions with strong excitation of the $\phi$-oscillator, remains
an open problem.

\paragraph*{Acknowledgements}
We wish to thank Boston University's Center for Computational Science and
Office of Information Technology for generous
allocations of supercomputer time.
The authors are indebted to A.~Barvinsky, F.~Bezrukov and A.~Vilenkin for helpful discussions.
This research was supported in part under DOE grant DE-FG02-91ER40676, 
Russian Foundaion for Basic Research grant 02-02-17398, 
Russian Council for Presidental Grants and State Support of Scientific Schools
grant 00-15-96626
and by the U.S. Civilian Research and Development Foundation 
for Independent States of FSU~(CRDF) award RP1-2103.
D.L.~gratefully ackowledges hospitality at Boston University 
and C.R.~at the Institute for Nuclear Research while 
this work was in progress.
\vspace{3mm}

\appendix
\section{Numerical techniques.}

\subsection{Classical solutions.}

To obtain the  solutions to eqs.~\eq{19*} and \eq{20*},
we chose the initial value {$b = -15$}, well outside the barrier, and
imposed the constraint \eq{20*} at the initial moment of time.
We found the classical solution by making use of
the fourth order Runge--Kutta method with adaptive 
step size (see, e.g., Refs.~\cite{Press}) for the first order
Hamiltonian equations of motion. We checked that the
constraint \eq{20*} was satisfied during entire evolution with relative
precision $10^{-9}$. For given initial oscillator phase $\theta_0$, 
the boundary
of the no-transition region of initial data was found with precision
$10^{-6}$ by ``division by 2'' search in $E_{b,0}$. The phase
$\theta_0$ was discretized into 100 points, and the minimum in $\theta_0$
of
the boundary of
the no-transition regions was found. 

\subsection{The discretized Wheeler--De~Witt equation.}

We solved the Wheeler--De~Witt equation \eq{11*} by expanding the
wave function in eigenstates of the oscillator of fixed frequency 
$\omega$,
\[
      \Psi (b) = \sum_n \tilde{C}_n (b) \Psi_n^{(\omega)}\;,
\]
where $\Psi_n^{(\omega)}$ 
are independent of $b$.\footnote{
This expansion, which is more convenient for computational 
purposes,  is related to the expansion into ``instantaneous'' modes of 
eq.~\eq{EqN1}, see A.5.} Asymptotically, $\tilde{C}_n$
coincide with the coefficients entering eq.~\eq{23*}, so the boundary
conditions are precisely \eq{24*}. The Wheeler--De~Witt equation 
takes the form of an ordinary differential equation for the vector
$\tilde{C}$,
\be
\left[ \Lambda \frac{d^2}{d b^2} 
+ A\right] {\tilde C}(b) = 0\;,
\label{14-dip}
\ee
where the matrix $A$ is defined as 
\begin{eqnarray}
A_{n n'}  =  &-& 2 \left[\frac{1}{\Lambda}\left(U(b) - \tilde\epsilon\right) - 
\omega(n + {1}/{2})\left(1 + \frac{M^2 f(b)}{2 \omega^2}\right)\right] 
\delta_{n',\;n} \nonumber \\
&+&\frac{M^2 f(b)}{2 \omega} \left(\sqrt{(n + 1)(n + 2)}\delta_{n',\; n + 2}
+ \sqrt{n(n - 1)} \delta_{n',\; n - 2}\right)\;.\nonumber
\label{15-dip}
\end{eqnarray}
To solve eq.~\eq{14-dip} numerically, we truncated the system to a finite
number of levels, i.e., set $0 \leq n \leq (N_0 -1)$. We considered the
system in a finite interval of $b$ and discretized
the variable $b$ by introducing a  lattice of spacing $\Delta$.  The
sites were chosen at points
\[
b_i = i \Delta\;, \label{L1}
\]
where $i = -N_b, \dots N_b$.

One way to discretize eq.~\eq{14-dip} would be to make use of the
symmetric approximation for the second derivative
\begin{eqnarray}
\frac{d^2 {\tilde C} (b)}{d b^2} = \frac{1}{\Delta^2}[{\tilde C}(b + 
\Delta) + {\tilde C}(b - \Delta) - 2 {\tilde C}(b)] + O(\Delta^4)\;.
\label{19-dip}
\end{eqnarray}
However, we made use of the Numerov--Cowling method
which has higher precision. Namely, the Taylor expansion of
$\tilde{C} (b)$ gives
\begin{eqnarray}
\tilde{C}_{i + 1} + \tilde{C}_{i - 1} - 2 \tilde{C}_i = 
\Delta^2 \left[\frac{d^2 \tilde{C}}{d b^2}\right]_i  + 
\frac{\Delta^4}{12} \left[\frac{d^4 \tilde{C}}{d b^4}\right]_i + 
O(\Delta^6)\;.\nonumber
\end{eqnarray}
Making use of eq.~\eq{14-dip}, one writes the second term on the right
hand side as
\begin{eqnarray}
-\frac{\Delta^4}{12 \Lambda} \frac{d^2}{d b^2}(A {\tilde C}) = 
- \frac{\Delta^2}{12 \Lambda} (A_{i + 1} {\tilde C}_{i + 1} + 
 A_{i - 1} 
{\tilde C}_{i - 1} - 2  A_i {\tilde C}_i) + O(\Delta^6)\;,
\end{eqnarray}
where eq.~\eq{19-dip} was used for the derivative of the vector
$(A \tilde{C})$. In this way we obtain the discretized
version of eq.~\eq{14-dip} with error $O(\Delta^6)$,
\begin{eqnarray}
\left(1 + \frac{\Delta^2}{12 \Lambda} A_{i - 1}\right) {\tilde C}_{i - 1}
+ \left(1 + \frac{\Delta^2}{12 \Lambda} A_{i + 1}\right){\tilde C}_{i + 1} - 
\left(2 - \frac{5 \Delta^2}{6 \Lambda} A_i\right){\tilde C}_i = 0.
\label{21-dip}
\end{eqnarray}
The unknowns of this finite difference equation are numbers
$\tilde{C}_{i,n}$ with {$i = -N_b, \dots N_b$, $n=0,\dots (N_0-1)$}.
So, the number of unknowns is
{$N_0(2N_b +1)$}. The system \eq{21-dip} is a set of $N_0(2N_b -1)$
equations. The boundary conditions \eq{24*}
after discretization become $2N_0$ linear equations
\begin{eqnarray}
{\tilde C}_{- N_b,\;n} - e^{i k_n \Delta} {\tilde C}_{- N_b + 1,\;n} & = & 
\delta_{n,\;n_0}\left(1 - e^{2 i k_n \Delta}\right)\;, \nonumber \\
{\tilde C}_{N_b - 1,\;n} - e^{- i k_n\Delta}{\tilde C}_{N_b,\;n} & = & 0\;.
\label{22-dip} 
\end{eqnarray}
Hence, the total number of equations coincides with the number of
unknowns, so $\tilde{C}_{i,n}$ are uniquely determined.

As we will see below, the solution of the discretized system well
approximates the solution of the original one only if the 
numbers of lattice points and oscillator levels are quite large, 
$N_b \sim 20000$,
$N_0 \sim 150$. Numerical methods for generic system of 
equations would fail in this case,
since the total number of complex equations is more than 3 million.
However, following Ref.~\cite{Bonini} we exploited the special form of
eq.~\eq{21-dip} 
to 
derive the solution by an iterative procedure.
Indeed, by inverting  numerically a set of $N_0 \times N_0$ 
matrices\footnote{An additional simplification is that the matrices
$(2-5A_i /6)$ that must  be inverted are sparse, with non-zero entries
on three diagonals only. Inverting  matrices of this type takes much
less computer time than with  to matrices of generic form.}, one
writes eq.~\eq{21-dip} in the following form,
\begin{eqnarray}
{\tilde C}_i = L_i {\tilde C}_{i - 1} + 
R_i {\tilde C}_{i + 1}\;,
\label{23-dip}
\end{eqnarray}
where $L_i$ and $R_i$ are again $N_0 \times N_0$
matrices. 
Equations~\eq{23-dip} can now be used to eliminate a set of 
$\tilde{C}_i$ variables with given $i$, leading to 
new equations, which, with some matrix algebra, 
can be cast again into the form~\eq{23-dip}.
Thus one can progressively eliminate all the variables
$\tilde{C}_i$ at the intermediate points
{${i = (- N_b + 1) \dots (N_b - 1)}$}
expressing them in terms of linear combinations of 
$\tilde{C}_{ - N_b}$ and $\tilde{C}_{N_b}$.
By substituting these linear combinations
into eq.~\eq{22-dip}, we obtained a system of linear inhomogeneous
equations for the complex vectors $\tilde{C}_{-N_b}$ 
and $\tilde{C}_{N_b}$ which was solved in straightforward way.
Then the actual values of  $\tilde{C}_{-N_b}$ 
and $\tilde{C}_{N_b}$  were introduced back into expressions for
the rest of unknowns, and in this way the complete wave function was
found.

It is worth pointing out that the procedure in which
intermediate variables are excluded involves operations
with real matrices only. Note also that the interaction between
different $\tilde{C}_n$ in eq.~\eq{14-dip} occurs only
if they have the same parity. Hence, it is sufficient to consider only
even (or only odd) modes.

The algorithm described above lends itself to parallel implementation:
the variables $\tilde{C}_i$ at odd sites are iliminated in parallel, 
and the procedure is repeated in a recursive manner. Our code scaled well 
with the number of processors $N_{\mbox{proc}}$, and most of the results
presented in this paper were obtained with $N_{\mbox{proc}} = 64$.

One must of course be careful that the effects of discretization
and truncation of the original
Wheeler-- De Witt equation are kept small. We used a value for the cut-off 
of the number of modes $N_0$ such that the
truncation energy exceeds the
height of the barrier by a factor of 2,
\[
N_0 = \left[\frac{2}{\omega \Lambda}\right]\;.\nonumber
\]
The lattice spacing was chosen equal to $1/12$ of the minumum
wavelength along $b$,
\[
\Delta = \frac{1}{12} \frac{2 \pi}{k_{N_0 - 1}}
= 
\frac{\sqrt{2}\pi\Lambda}{12 \sqrt{2 + \tilde{\epsilon}}} 
\;.\nonumber
\]
We used the lattice of the size
\[
N_b = \frac{15}{\Delta}\;,
\]
so that the potential is very small near the end points.
We performed various checks, summarized in Appendix B, which show that
this choice of parameters is appropriate.

\subsection{Iterative improvement of precision.}

In our calculation we used a double precision representation of
real numbers, which entails round-off errors of order $10^{-15}$
to $10^{-16}$.  The exponential behavior in $b$ and $n$ gives
origin to numbers $\tilde C_{i,n}$ varying over a range
that extends well beyond the limits of double precision.  What
makes it still possible for us to use double precision 
is that only coefficients $\tilde C_{i,n}$ with neighboring 
values of $i$ and $n$  and comparable magnitude are coupled by the
equations that must be solved.  Thus one
never encounters additions or subtractions of variables with
wildly differing exponents.  The elimination procedure used
for solving the equations can nevertheless propagate round-off
errors, so that special care is needed to ensure that the
values of the smaller $\tilde C_{i,n}$ are accurate. We 
accomplished this by an iterative improvement of precision,
which we briefly describe here.  

Let $\tilde C^{(0)}_{i,n}$
be the initial solution obtained by the elimination procedure.
The propagation of round-off errors will give origin to
absolute errors of order $\epsilon = 10^{-13}$ to $10^{-14}$ 
in the coefficients $\tilde C_{i,n}$. Thus those
coefficients whose true value is smaller than $\epsilon $ will
turn out to be totally wrong and the equations coupling
them will have errors of the same order of magnitude.
It is convenient to denote these equations symbolically
by $\hat A\, C=0$.  Substituting into the equations the numbers
$C^{(0)}$ we obtain instead $\hat A \,  C^{(0)}=B^{(0)}$, where
$B^{(0)} \sim \epsilon $.  We introduce
therefore a correction $\delta C^{(0)}$, which we find
by solving $\hat A \,\, \delta C^{(0)}= - B^{(0)}$. Note that the equations for 
$\delta C^{(0)}$ only involve right-hand side  terms of order $\epsilon $,
so that the propagation of round-off errors will now give origin
to errors of much smaller absolute magnitude $\sim \epsilon^2 $.
With $\delta C^{(0)}$ we form the first iterate of the solution 
$C^{(1)}=C^{(0)}+\delta C^{(0)}$.  While the numbers $C^{(0)}$
and $\delta C^{(0)}$ will typically not be smaller than $\epsilon $,
the first iterates $C^{(1)}$ corresponding to coefficients whose
true value is less than $\epsilon $ will accordingly take
smaller values, being now affected by errors
$\sim \epsilon^2$. Inserting the first iterate $C^{(1)}$ into 
the equations we find now errors $B^{(1)} = \hat A \,  C^{(1)}\sim \epsilon^2$
and the procedure can be repeated until the values of the smallest
coefficients become stable.

\begin{figure}[ht!]
\begingroup%
  \makeatletter%
  \newcommand{\GNUPLOTspecial}{%
    \@sanitize\catcode`\%=14\relax\special}%
  \setlength{\unitlength}{0.1bp}%
\begin{picture}(3600,2160)(0,0)%
\special{psfile=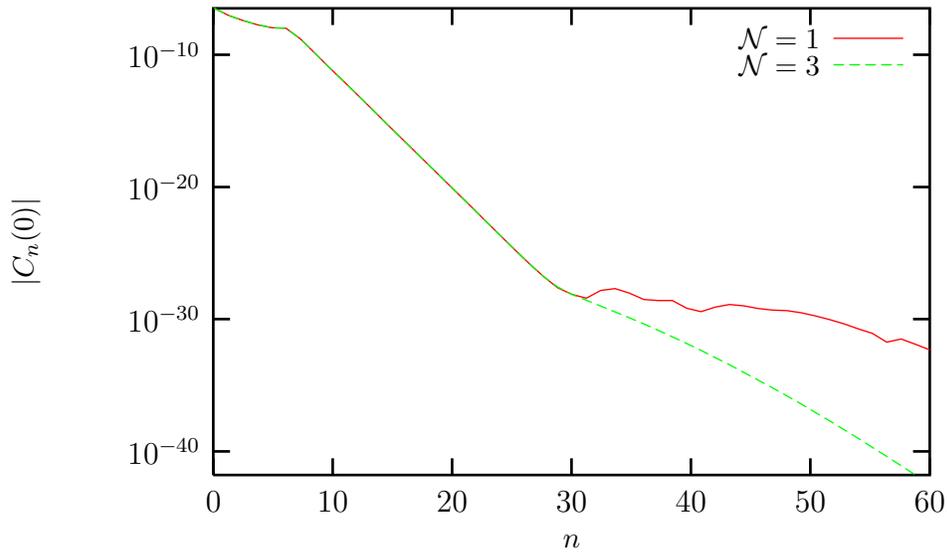 llx=0 lly=0 urx=720 ury=504 rwi=7200}
\put(3037,1847){\makebox(0,0)[r]{\small ${\cal N} = 3$}}%
\put(3037,1947){\makebox(0,0)[r]{\small ${\cal N} = 1$}}%
\put(2100,50){\makebox(0,0){\small $n$}}%
\put(100,1180){%
\special{ps: gsave currentpoint currentpoint translate
270 rotate neg exch neg exch translate}%
\makebox(0,0)[b]{\shortstack{\small $|C_n (0)|$}}%
\special{ps: currentpoint grestore moveto}%
}%
\put(3450,200){\makebox(0,0){60}}%
\put(3000,200){\makebox(0,0){50}}%
\put(2550,200){\makebox(0,0){40}}%
\put(2100,200){\makebox(0,0){30}}%
\put(1650,200){\makebox(0,0){20}}%
\put(1200,200){\makebox(0,0){10}}%
\put(750,200){\makebox(0,0){0}}%
\put(700,389){\makebox(0,0)[r]{$10^{-40}$}}%
\put(700,888){\makebox(0,0)[r]{$10^{-30}$}}%
\put(700,1386){\makebox(0,0)[r]{$10^{-20}$}}%
\put(700,1885){\makebox(0,0)[r]{$10^{-10}$}}%
\end{picture}%
\endgroup
 
\caption{\label{fig10} Wave function at $b=0$ at different steps
of the iteration procedure.}
\end{figure}
\begin{figure}[ht!]
\begingroup%
  \makeatletter%
  \newcommand{\GNUPLOTspecial}{%
    \@sanitize\catcode`\%=14\relax\special}%
  \setlength{\unitlength}{0.1bp}%
\begin{picture}(3600,2160)(0,0)%
\special{psfile=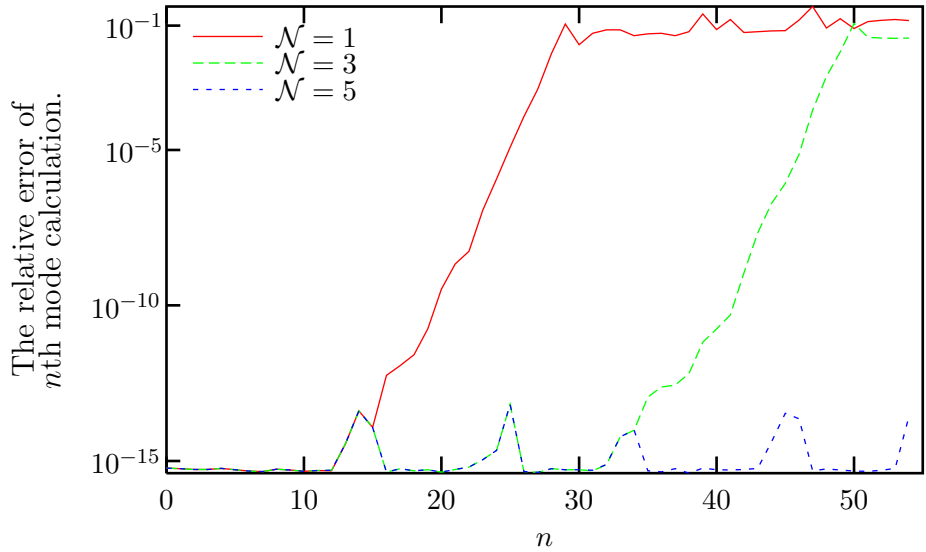 llx=0 lly=0 urx=720 ury=504 rwi=7200}
\put(1013,1747){\makebox(0,0)[l]{\small ${\cal N} = 5$}}%
\put(1013,1847){\makebox(0,0)[l]{\small ${\cal N} = 3$}}%
\put(1013,1947){\makebox(0,0)[l]{\small ${\cal N} = 1$}}%
\put(315,346){\makebox(0,0)[l]{\small $10^{-15}$}}%
\put(315,933){\makebox(0,0)[l]{\small $10^{-10}$}}%
\put(367,1519){\makebox(0,0)[l]{\small $10^{-5}$}}%
\put(367,1988){\makebox(0,0)[l]{\small $10^{-1}$}}%
\put(2025,50){\makebox(0,0){\small $n$}}%
\put(200,1180){%
\special{ps: gsave currentpoint currentpoint translate
270 rotate neg exch neg exch translate}%
\makebox(0,0)[b]{\shortstack{ $n$th mode calculation.}}%
\special{ps: currentpoint grestore moveto}%
}%
\put(100,1180){%
\special{ps: gsave currentpoint currentpoint translate
270 rotate neg exch neg exch translate}%
\makebox(0,0)[b]{\shortstack{The relative error of }}%
\special{ps: currentpoint grestore moveto}%
}%
\put(3191,200){\makebox(0,0){\small $50$}}%
\put(2673,200){\makebox(0,0){\small $40$}}%
\put(2155,200){\makebox(0,0){\small $30$}}%
\put(1636,200){\makebox(0,0){\small $20$}}%
\put(1118,200){\makebox(0,0){\small $10$}}%
\put(600,200){\makebox(0,0){\small $0$}}%
\end{picture}%
\endgroup
 
\caption{\label{fig11} Relative error at different steps
of the iteration procedure.}
\end{figure}

We show in Fig.~\ref{fig11} 
the relative error as function of $n$ after
different numbers of iterations. This quantity is defined as the ratio
of the modulus of the left hand side of eq.~\eq{23-dip} to the sum of
moduli of all terms in this equation. The absolute value of
$\tilde{C}_{i,n}$ at $b=0$ at different steps of the iteration
procedure is presented in Fig.~\ref{fig10}. We conclude that the iteration
procedure indeed enables one to obtain the solutions with relative, 
not absolute,
error of order $10^{-13}$.

\subsection{Total computational time.}

Our main emphasis in this paper is the study of the solutions
to the Wheeler--De~Witt equation at small $\Lambda$. Let us estimate
the dependence of the time of calculations on this parameter.
At each step of the iteration procedure of section A.3, 
$N_b\cdot N_0^3$ floating point operations are performed.
The number of iterations is proportional to logarithm of the minimum
value of the wave function. The latter scales as 
$N_{\mbox{iter}} \propto 1/\Lambda$, see 
eq.~\eq{35*}. Thus, the time of calculations is proportional to
\[
     N_b\cdot N_0^3 \cdot N_{\mbox{iter}} \propto \frac{1}{\Lambda^5}\;.
\]
This estimate shows that reaching lower values of $\Lambda$
occurs with dramatic increase of processor time.
 
In practice, for $\Lambda = 0.0225$ and $\omega = 0.6$,
the time of calculations with 64 processors was equal to
30 to 40 minutes. At larger values of $\omega$ the time of
calculations was somewhat smaller.

\subsection{Bogoliubov transformation.}
 
Numerical calculations were conveniently performed in the basis
of the eigenfunctions of the oscillator of $b$-independent
frequency $\omega$. The interpretation of the results is done, however, 
in the basis of the ``instantaneous'' Hamiltonian, describing
the oscillator of frequency $\Omega(b)$.  So, for every $b$
we performed
the Bogolubov transformation and obtained the coefficients $C_n$
starting from~$\tilde{C}_n$.

The key element of this procedure is a set of transformation 
coefficients
\[
 \langle n_{\Omega} \vert m_{\omega} \rangle
 = \langle 0_{\Omega}\vert \frac{\alpha_\Omega^n}{\sqrt{n!}}  
\vert m_\omega \rangle
=  \frac{1}{\sqrt{n!}}
\sum\limits_{k=0}^{2 N_0 -1}\langle 0_\Omega \vert k_\omega \rangle
\langle k_\omega \vert \alpha_\Omega^n \vert m_\omega \rangle;\;\;n,\;
m < N_0\;. 
\]
Here $\alpha_{\Omega}$ denotes the annihilation operator in the basis
of the ``instantaneous'' Hamiltonian. The first factor on the right
hand side is given explicitly by 
\begin{eqnarray}
\langle 0_\Omega| 2 l_\omega \rangle = (\omega \Omega)^{1/4} 
\sqrt{\frac{2}{\omega + \Omega}} \frac{(2 l)!}{2^l \; l!} 
\left(\frac{\omega - \Omega}{\omega + \Omega}\right)^l\;, \nonumber
\end{eqnarray} 
while the matrix elements 
$\langle k_\omega \vert \alpha_\Omega^n \vert m_\omega \rangle$
can be calculated numerically, using the harmonic oscillator algebra.
 
\vspace{3mm}

\section{Checks of numerical procedure.}

Since we were dealing with very small values of the coefficients
$\tilde{C}_{i,n}$, we had to perform a series of checks of the entire
procedure. To make sure that round-off errors do not
spoil our calculations, we repeated the calculations
with quadruple precision (32 decimal places), and found that the
results
for the wave function coincided with those obtained with double
precision, with relative error $10^{-13}$. This check was performed for
the interaction function $f_3$ and the following values of parameters:
$\sqrt{\Lambda} = 0.3$, $0.21$ and $0.17$; $M =0.2$; $\omega = 0.6$;
$\tilde{\epsilon}= 0.27$.

We also checked that the effects of discretization of $b$ and
truncation of the oscillator energy levels are small, by varying the
parameters $N_b$, $N_0$ and $\Delta$ by a factor of 2. The interaction
function in this check was $f_3$, the parameters were
$\sqrt{\Lambda} = 0.21$; $M=0.2$; $\omega =0.6$; $\tilde{\epsilon}=
0.27$. The relative difference between the solutions was always less
than $10^{-3}$.

Another way to check our procedure is to make use of the current
\begin{eqnarray}
J = \frac{1}{i} \left({\tilde C}^+ \partial_b {\tilde C} - 
\partial_b {\tilde C}^+ {\tilde C}\right),\;\nonumber
\end{eqnarray}  
which is conserved in the continuum theory as a consequence
of the Wheeler--De~Witt equation, i.e.,
\[
      \partial_b J = 0\;,
\]
while 
the discretized version does not have this property. By calculating
the value of the current for every lattice site, we checked the
discretization error. The precision of the calculation of $J$ for
given set of vectors $\tilde{C}_i$ must be $O(\Delta^6)$,
otherwise the conservation of current would hold with lower precision
than the precision the solution was found. From the Taylor expansion
 we find
\begin{eqnarray}
J_i & = & \mbox{Im} 
\left\{ \displaystyle\frac{7 \Delta^3}{180} 
     {\tilde C}_i^+  \frac{A_i^2}{\Lambda^2} {\tilde C}_{i + 1} + 
\displaystyle\frac{\Delta^3}{72}
 {\tilde C}_i^+ \frac{A_i}{\Lambda} \frac{A_{i + 1}}{\Lambda} 
{\tilde C}_{i + 1}  +  
  \displaystyle\frac{\Delta^4}{36}{\tilde C}_i^+ 
\frac{(\partial_b A)_i}{\Lambda} 
\frac{A_i}{\Lambda} {\tilde C}_{i + 1}
 - \right. \nonumber\\
&&
 \left. \displaystyle\frac{\Delta^4}{30} {\tilde C}_i^+ 
\frac{(\partial_b A)_i}{\Lambda} \frac{A_i}{\Lambda} {\tilde {C}_i}
  + \displaystyle\frac{\Delta^5}{40}{\tilde C}_i^+
\frac{(\partial_b^2 A)_i}{\Lambda} \frac{A_i}{\Lambda} {\tilde {C}_i}
  + \displaystyle\frac{\Delta^3}{72} {\tilde C}_i^+ \frac{A_i}{\Lambda}
\frac{A_{i - 1}}{\Lambda}
{\tilde C}_{i - 1}\right. 
 + \nonumber\\
&&
 \left.\displaystyle\frac{\Delta^3}{20} {\tilde C}_i^+ 
\frac{(\partial_b^2 A)_i}{\Lambda} 
 {\tilde C}_{i + 1} - 
 \displaystyle\frac{\Delta}{90} {\tilde C}_i^+ \frac{A_{i + 1}}{\Lambda} 
{\tilde C}_{i + 1}  
  + 
\displaystyle\frac{\Delta}{3}{\tilde C}_i^+ \frac{A_i}{\Lambda} {\tilde C}_{i + 1}  
  \label{26-dip} + 
\right.\\
&&
\left.\displaystyle\frac{ \Delta^2}{6} 
{\tilde C}_i^+ \frac{(\partial_b A)_i}{\Lambda} {\tilde C}_{i + 1}
  + 
 \displaystyle\frac{\Delta}{360} {\tilde C}_i^+ \frac{A_{i + 2}}{\Lambda} 
{\tilde C}_{i + 2} 
  + \displaystyle\frac{\Delta}{360} {\tilde C}_i^+ \frac{A_{i - 2}}{\Lambda} 
{\tilde C}_{i - 2} - 
 \right.\nonumber \\ 
&&
\left.  \displaystyle\frac{\Delta}{90} 
{\tilde C}_i^+ \frac{A_{i - 1}}{\Lambda} {\tilde C}_{i - 1}
  + \displaystyle \frac{2}{\Delta}{\tilde C}_i^+ {\tilde C}_{i + 1} 
\right\}
  + O(\Delta^6)\nonumber\;.
 \end{eqnarray} 
Here we made use of eq.~\eq{14-dip} to express higher derivatives of
  $\tilde{C}_i$. The first and second derivatives of the matrix
$A$ were found by straightforward differentiation of the
  expression \eq{15-dip}.

Making use of the calculated current $J$, we estimated the relative
error
of the calculation of $\tilde{C}_i$,
\begin{eqnarray}
\delta {\tilde C}_i =  \frac{|J - J_{exact}|}{k_{N_0 - 1} |{\tilde C}_i|^2}\;, 
\label{27-dip}
\end{eqnarray}
where the exact value of the current was found in the
asyptotic region of large positive $b$,
\begin{eqnarray}
J_{exact} = \lim\limits_{b \to \infty} 
\left( \sum\limits_{n = 0}^{\infty} 2 k_n |{\tilde C}_{n}(b)|^2\right) 
\approx \sum\limits_{n = 0}^{N_0 - 1} 2 k_n |{\tilde C}_{N_b,\;n}|^2. 
\label{28-dip}
\end{eqnarray}
Because of the exponential suppression of the wave function behind the
barrier, eq.~\eq{28-dip} is satisfied with better precision, as compared to
the precision at which the current is calculated in  
 the classically
forbidden region. Equations \eq{26-dip}, \eq{27-dip} and \eq{28-dip} 
give an
estimate of the relative error in the calculation of
$\tilde{C}_{i,n}$, which we found to be
less than $10^{-7}$.
 
As one more check we changed the sign of the Hamiltonian of the scale
factor, and obtained a conventional quantum mechanical system of two
degrees of freedom. We checked for that system,
that the sum of reflection and
transmission coefficients was equal to 1 with precision $10^{-10}$.

Finally, we compared the numerical solutions  at small $M$ and
$\Lambda$ to the analytic  solutions obtained within the
adiabatic approximation of section 2.2.  The first two modes of these
two solutions coincided
within 3 \%. The latter, fairly large deviation we ascribe to
the approximations involved in obtaining
the analytic solutions.


\end{document}